\documentclass[twocolumn,showpacs,nofootinbib,superscriptaddress,amsmath,amssymb,aps]{revtex4-2}
 
 \usepackage[pdftex]{graphicx} 

 \usepackage{textcomp} 

 \usepackage{cancel} 

 \usepackage[usenames,svgnames,table]{xcolor} 

 \usepackage[english]{babel} 

 \usepackage{ucs} 

 \usepackage[utf8]{inputenc} 

 \usepackage{siunitx} 

 \usepackage[version=3]{mhchem} 

 \usepackage{multirow} 

 \usepackage{soul} 

 \usepackage{pslatex} 

 \usepackage{dcolumn}

 \usepackage{bm}

 \usepackage{mciteplus} 

\newcommand{\hf}{\textsc{HF$\ell$}} 

\newcommand{\cm}{c.m.}

\newcommand{\ca}{$^{12}$C}

\newcommand{\solf}{$^{32}$S}

\newcommand{\neo}{$^{20}$Ne}

\usepackage[pdftex]{hyperref}
\hypersetup{
    colorlinks=true,
    linkcolor=blue,
    urlcolor=blue,
    citecolor=blue
}

\begin{document} 

%


%


\title{Examination of cluster production in excited light systems at Fermi  energies from new experimental data and comparison with transport model calculations}

\author{C.~Frosin}  \email{catalin.frosin@unifi.it} 
 \affiliation{Dipartimento di Fisica, Universit\`a di Firenze, I-50019 Sesto Fiorentino, Italy} 
 \affiliation{INFN Sezione di Firenze, I-50019 Sesto Fiorentino, Italy} 

\author{S.~Piantelli} 
 \affiliation{INFN Sezione di Firenze, I-50019 Sesto Fiorentino, Italy} 

\author{G.~Casini} 
 \affiliation{INFN Sezione di Firenze, I-50019 Sesto Fiorentino, Italy} 
 
\author{A.~Ono} 
 \affiliation{Department of Physics, Tohoku University, Sendai 980-8578, Japan}   

\author{A.~Camaiani} 
 \affiliation{Instituut voor Kern- en Stralingsfysica, K.U. Leuven, Celestijnenlaan 200D, B-3001 Leuven, Belgium}
  
\author{L.~Baldesi}  
 \affiliation{INFN Sezione di Firenze, I-50019 Sesto Fiorentino, Italy} 
    
\author{S.~Barlini}  
 \affiliation{Dipartimento di Fisica,  Universit\`a di Firenze, I-50019 Sesto Fiorentino, Italy} 
 \affiliation{INFN Sezione di Firenze, I-50019 Sesto Fiorentino, Italy} 
 
\author{B.~Borderie}
 \affiliation{Université Paris-Saclay, CNRS/IN2P3, IJCLab, 91405 Orsay, France}

\author{R.~Bougault} 
 \affiliation{Normandie Université, ENSICAEN, UNICAEN, CNRS/IN2P3, LPC Caen, 14000 Caen, France} 

\author{C.~Ciampi}
 \affiliation{Dipartimento di Fisica,  Universit\`a di Firenze, I-50019 Sesto Fiorentino, Italy} 
 \affiliation{INFN Sezione di Firenze, I-50019 Sesto Fiorentino, Italy}

\author{M.~Cicerchia} 
 \affiliation{INFN Laboratori Nazionali di Legnaro, 35020 Legnaro, Italy} 

\author{A.~Chbihi} 
 \affiliation{Grand Accélérateur National d’Ions Lourds (GANIL), CEA/DRF–CNRS/IN2P3, Boulevard Henri Becquerel, F-14076 Caen, France} 
 
\author{D.~Dell’Aquila} 
 \affiliation{Dipartimento di Scienze Chimiche, Fisiche, Matematiche e Naturali, Università degli Studi di Sassari, 07100 Sassari Italy } 
 \affiliation{INFN Laboratori Nazionali del Sud, 95123 Catania, Italy}

\author{J.~A.~Dueñas} 
 \affiliation{Departamento de Ingeniería Eléctrica y Centro de Estudios Avanzados en Física, Matemáticas y Computación,
Universidad de Huelva, 21007 Huelva, Spain} 

\author{D.~Fabris} 
 \affiliation{INFN Sezione di Padova, 35131 Padova, Italy} 
 
\author{Q.~Fable} 
 \affiliation{Laboratoire des 2 Infinis - Toulouse (L2IT-IN2P3), Universit\'e de Toulouse, CNRS, UPS, F-31062 Toulouse Cedex 9, France}  

\author{J.~D.~Frankland} 
 \affiliation{Grand Accélérateur National d’Ions Lourds (GANIL), CEA/DRF–CNRS/IN2P3, Boulevard Henri Becquerel, F-14076 Caen, France} 
 
\author{T.~Génard} 
 \affiliation{Grand Accélérateur National d’Ions Lourds (GANIL), CEA/DRF–CNRS/IN2P3, Boulevard Henri Becquerel, F-14076 Caen, France}  
 
\author{F.~Gramegna} 
 \affiliation{INFN Laboratori Nazionali di Legnaro, 35020 Legnaro, Italy}    

\author{D.~Gruyer} 
 \affiliation{Normandie Université, ENSICAEN, UNICAEN, CNRS/IN2P3, LPC Caen, 14000 Caen, France} 

\author{M.~Henri} 
 \affiliation{Grand Accélérateur National d’Ions Lourds (GANIL), CEA/DRF–CNRS/IN2P3, Boulevard Henri Becquerel, F-14076 Caen, France} 

\author{B.~Hong} 
 \affiliation{ Center for Extreme Nuclear Matters (CENuM), Korea University, Seoul 02841, Republic of Korea}
 \affiliation{ Department of Physics, Korea University, Seoul 02841, Republic of Korea}

\author{M.~J.~Kweon} 
 \affiliation{Department of Physics, Inha University, Incheon 22212, Republic of Korea}
    
\author{S.~Kim} 
 \affiliation{ Institute for Basic Science, Daejeon 34126, Republic of Korea}

\author{A.~Kordyasz}
 \affiliation{Heavy Ion Laboratory, University of Warsaw, 02-093 Warszawa, Poland}
    
\author{T.~Kozik} 
 \affiliation{Faculty of Physics, Astronomy and Applied Computer Science, Jagiellonian University, 30-348 Kracow, Poland} 
    
\author{I.~Lombardo} 
 \affiliation{INFN Sezione di Catania, 95123 Catania, Italy} 
 \affiliation{Dip. di Fisica e Astronomia, Università di Catania, via S. Sofia 64, 95123 Catania, Italy}
 
\author{O.~Lopez} 
 \affiliation{Normandie Université, ENSICAEN, UNICAEN, CNRS/IN2P3, LPC Caen, 14000 Caen, France}
 
\author{T.~Marchi} 
 \affiliation{INFN Laboratori Nazionali di Legnaro, 35020 Legnaro, Italy} 
 
\author{K.~Mazurek} 
 \affiliation{Institute of Nuclear Physics Polish Academy of Sciences, Radzikowskiego 152, Krakow, PL-31342, Poland} 
 
\author{S.~H.~Nam} 
 \affiliation{ Center for Extreme Nuclear Matters (CENuM), Korea University, Seoul 02841, Republic of Korea}
 \affiliation{ Department of Physics, Korea University, Seoul 02841, Republic of Korea}
  
\author{J.~Lemarié} 
 \affiliation{Grand Accélérateur National d’Ions Lourds (GANIL), CEA/DRF–CNRS/IN2P3, Boulevard Henri Becquerel, F-14076 Caen, France}  
 
\author{N.~LeNeindre} 
 \affiliation{Normandie Université, ENSICAEN, UNICAEN, CNRS/IN2P3, LPC Caen, 14000 Caen, France} 
 
\author{P.~Ottanelli} 
 \affiliation{INFN Sezione di Firenze, I-50019 Sesto Fiorentino, Italy}

\author{M.~Parlog} 
 \affiliation{Normandie Université, ENSICAEN, UNICAEN, CNRS/IN2P3, LPC Caen, 14000 Caen, France}
 \affiliation{“Horia Hulubei” National Institute of Physics and Nuclear Engineering (IFIN-HH), RO-077125 Bucharest Magurele, Romania}

\author{J.~Park} 
 \affiliation{ Center for Extreme Nuclear Matters (CENuM), Korea University, Seoul 02841, Republic of Korea}
 \affiliation{ Department of Physics, Korea University, Seoul 02841, Republic of Korea}

\author{G.~Pasquali} 
 \affiliation{Dipartimento di Fisica, Universit\`a di Firenze, I-50019 Sesto Fiorentino, Italy} 
 \affiliation{INFN Sezione di Firenze, I-50019 Sesto Fiorentino, Italy}

\author{G.~Poggi} 
 \affiliation{Dipartimento di Fisica, Universit\`a di Firenze, I-50019 Sesto Fiorentino, Italy} 
 \affiliation{INFN Sezione di Firenze, I-50019 Sesto Fiorentino, Italy}

\author{A.~Rebillard-Soulié} 
 \affiliation{Normandie Université, ENSICAEN, UNICAEN, CNRS/IN2P3, LPC Caen, 14000 Caen, France} 
  
\author{B.~H.~Sun} 
 \affiliation{School of Physics and Nuclear Energy Engineering, Beihang University, Beijing 100191, China}
  
\author{A.~A.~Stefanini} 
 \affiliation{Dipartimento di Fisica, Universit\`a di Firenze, I-50019 Sesto Fiorentino, Italy} 
 \affiliation{INFN Sezione di Firenze, I-50019 Sesto Fiorentino, Italy}
  
\author{S.~Terashima} 
 \affiliation{School of Physics and Nuclear Energy Engineering, Beihang University, Beijing 100191, China}
 
\author{S.~Upadhyaya} 
 \affiliation{Faculty of Physics, Astronomy and Applied Computer Science, Jagiellonian University, 30-348 Kracow, Poland} 
 
\author{S.~Valdré} 
 \affiliation{INFN Sezione di Firenze, I-50019 Sesto Fiorentino, Italy}
 
\author{G.~Verde}
 \affiliation{INFN Sezione di Catania, 95123 Catania, Italy}
 \affiliation{Laboratoire des 2 Infinis - Toulouse (L2IT-IN2P3), Universit\'e de Toulouse, CNRS, UPS, F-31062 Toulouse Cedex 9, France}

\author{E.~Vient} 
 \affiliation{Normandie Université, ENSICAEN, UNICAEN, CNRS/IN2P3, LPC Caen, 14000 Caen, France}

\author{M.~Vigilante} 
 \affiliation{Dipartimento di Fisica, Università di Napoli, 80126 Napoli, Italy}
 \affiliation{INFN Sezione di Napoli, 80126 Napoli, Italy}
 
\collaboration{INDRA-FAZIA Collaboration}
\noaffiliation
   

\begin{abstract} 
Four different reactions, $^{32}$S+$^{12}$C and $^{20}$Ne+$^{12}$C at 25 and 50 MeV/nucleon, have been measured with the FAZIA detector capable of full isotopic identification of most forward emitted reaction products. Fragment multiplicities, angular distributions and energy spectra have been measured and compared with Monte Carlo simulations, i.e. the antisymmetrized molecular dynamics (AMD) and the heavy-ion phase space exploration (HIPSE) models. These models are combined with two different afterburner codes (\hf \:and SIMON) to describe the decay of the excited primary fragments. In the case of AMD, the effect of including the clustering and \textcolor{black}{inter-clustering processes} to form bound particles and fragments is discussed. A clear confirmation of the role of the cluster aggregation in the reaction dynamics and particle production for these light systems, for which the importance of the clustering process increases with bombarding energy, is obtained. 

\end{abstract} 




\maketitle

\section{Introduction} 

In the Fermi energy regime (20-100 MeV/nucleon), a wide range of phenomena, governed by the competition of the nuclear mean-field and the increasing role of nucleon-nucleon collisions, takes place. The associated phenomenology is very broad and includes many open channels in both peripheral and central reactions. Therefore, it is quite a challenge to develop models that accurately describe the entire range of the Fermi energies and impact parameter. Moreover, the inclusion of clustering effects and their influence on nuclear dynamics is still one of the open questions in nuclear physics~\cite{Beck_2017,ONO2019139,Guo,ebran2014,girod2013,borderie2016,borderie2021}. Indeed, there are ample evidence that clustering degrees of freedom play an important role in nuclear reactions~\cite{Freer_1997,Morelli_PhysRevC.99, Morelli_2014_v2,Baiocco054614,TIAN_PhysRevC.95.044613,Tian2_PhysRevC.97.034610, HanPhysRevC.102.064617}. 
In experiments, fragments and clusters are abundantly produced in various reactions, such as in the neck or mid-velocity component of dissipative binary reactions, in central collisions during the expansion phase of the system, and/or in semi-peripheral collisions where the excited quasi-projectile (henceforth referred to as QP) fragment breaks into smaller pieces~\cite{PiantelliPhysRevLett.88.052701,LEFORT2000397,REISDORF2010366,HagelPhysRevC.50.2017}.
In most of these cases, only a small percentage ($\approx$10\% at low energies and $\approx$30\% above 100 MeV/nucleon) of the total protons in the system is emitted as isolated particles. The various types of nuclear reactions represent an important tool to study cluster effects, and conversely, the cluster production is needed to interpret the reaction mechanisms, especially when lighter systems (\textit{A}$_{tot}<$50) are involved, as those of this paper.  

While experimental data for light systems are abundant in the Coulomb barrier energy region~\cite{Freer_1997, Freer_2007, Baiocco054614, Morelli_PhysRevC.99, Morelli_2014_v2, CamaianiPhysRevC.97.044607},
data in the Fermi energy region are rather sparse, either collected in the systematics of fusion~\cite{Eudes_PhysRevC.90.034609} (thus basically oriented to central collisions) or published in studies on more peripheral collisions as in~\cite{Glasow1990, TARASOV2004536, DAYRAS1986299}. 

The interest in light ion reactions at Fermi energy and beyond has also been renewed due to hadrontherapy, where the physical dose deposition is significantly affected by the inelastic interactions and fragmentation of ions along the penetration path in human tissues. In these cases, the ability of appropriate Monte Carlo codes to reproduce the differential yield of charged fragments is fundamental for treatment planning~\cite{Dudouet1,Dudouet2, DeNapoli2012}. 

Many models have been developed to describe the processes involved in the collision of heavy ions and observed in experimental data (see Refs.~\cite{Zhang, ONO2019139} for detailed reviews). Among these, microscopic transport models such as the antisymmetrized molecular dynamics (AMD) model of Ono et al.~\cite{Ono1,Ono2,Ono3} have achieved great success in describing many nuclear reaction phenomena for intermediate energy heavy-ion collisions~\cite{Camaiani_PhysRevC.102.044607,PiantelliPhysRevC.101.034613,Camaini_PhysRevC.103.014605,MockoPhysRevC.78.024612,Ono4, Takemoto_PhysRevC.57.811,Takemoto_PhysRevC.54.266, PiantelliPhysRevC.99.064616}. 
As shown in previous works~\cite{TIAN_PhysRevC.95.044613,Tian2_PhysRevC.97.034610,HanPhysRevC.102.064617}, the nucleon or complex particle correlations to aggregate together has a significant impact on the overall collision dynamics, for the formation of both light clusters and intermediate-mass fragments (IMFs)\footnote{here IMF refers to fragments from lithium to boron}. 
The latest version of AMD can handle these particle correlations in a stochastic way~\cite{ONO2019139, Ono4, Ono5, OnoJPSCP.32.010076} incorporating these effects in a well theoretically based framework.

This paper investigates the role of clusters through a comparison among a set of new experimental data obtained with the FAZIA apparatus~\cite{Bougault2014} and model calculations and verifies the importance of clustering in collisions of two light \textit{N}=\textit{Z} nuclei. We examine the effects of clustering in the \solf{}+\ca{} and \neo{}+\ca{} reactions at 25 and 50 MeV/nucleon using two different versions of the AMD model: one version of AMD without cluster correlations and one including them. So far, this type of studies has been performed mainly on C+C systems~\cite{TIAN_PhysRevC.95.044613,Tian2_PhysRevC.97.034610,HanPhysRevC.102.064617} and this paper represents a further test for these calculations extended to heavier systems.

As an additional benchmark, we use our new data for a comparison with the predictions of the heavy-ion space exploration (HIPSE) event generator~\cite{Lacroix_PhysRevC.69.054604}. HIPSE is a model meant to reproduce several features of Fermi energy reactions and in particular the pre-equilibrium and fast emission at midvelocities; indeed it takes into account nucleon-nucleon collisions as a crucial ingredient, assumed to occur with a probability to be tuned through comparison with experimental data. Although HIPSE treats nucleon collisions in a simplified manner, it also includes the formation of clusters which is relevant for the present case. The HIPSE generator has demonstrated to be a relatively fast and reliable reaction simulator from light to heavy systems~\cite{Vient_PhysRevC.98.044612,Dudouet2}, therefore it will be also compared with the new present FAZIA data.

The article is organized as follows. Section~\ref{sec:exp} describes the experimental apparatus and the data acquisition. Then, in section~\ref{sec:Models}, the main features of the Monte Carlo models are discussed. Section~\ref{res} reports the experimental results and the model predictions and discusses them starting from the general reaction characteristics and inclusive observables then focusing on more exclusive quantities. Conclusions are drawn in Section~\ref{sec:concl}.

\section{\label{sec:exp}Experimental Apparatus} 

\begin{figure}[t] 
   \centering 
   \includegraphics[width=1.0\columnwidth]{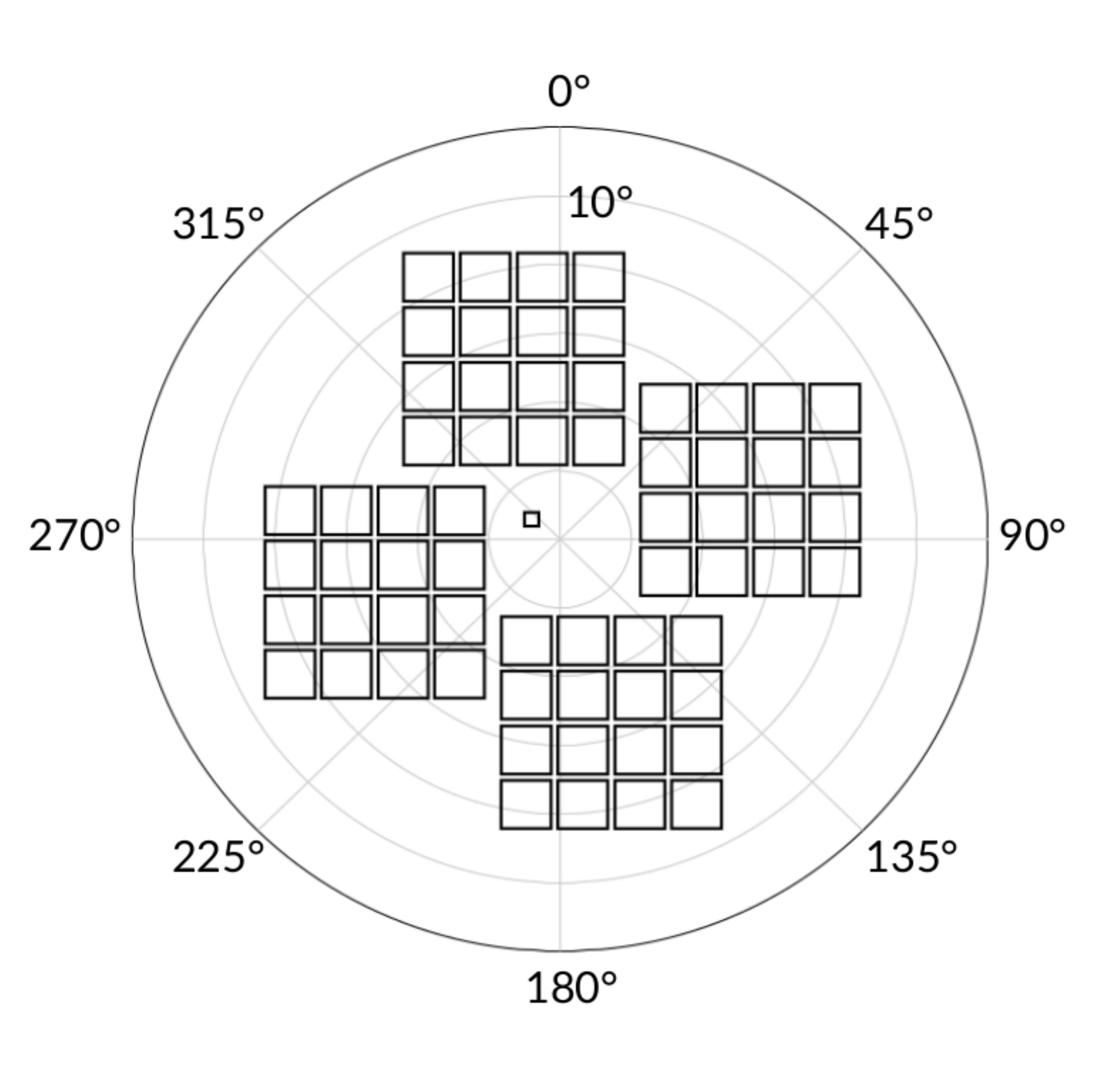}
   \caption{Polar view of the four FAZIA blocks in the geometry of the experiment. The beam axis passes through the center. The small detector close to the the center represents a beam monitor telescope. Within the precision of the drawing the geometrical square shape of the detectors is not deformed in the polar representation taking into account the very forward angles.} 
   \label{fig:GeoFazia} 
\end{figure}

In this experiment, labelled as FAZIA-Cor hereafter, the setup consisted of 4 blocks of the FAZIA array. Here, we briefly describe the main features and operating principles of the FAZIA detectors; these detectors are described in more detail in its parts and/or in its entirety in Refs.~\cite{Bougault2014, BARDELLI2009353, BARLINI2009644,LENEINDRE2013145, BARDELLI2011272, PASTORE201742,CARBONI2012251,FROSIN2020163018, VALDRE201927,parlog2010,hamrita2011}. 
The FAZIA-Cor experiment was performed with pulsed beams of \solf{} and \neo{} at 25 and 50 MeV/nucleon delivered by the superconducting cyclotron of INFN-LNS at typical currents of about 10$^9$~pps. The beams impinged on an enriched \ca{} target with a thickness of 240 $\mu$g/cm$^2$. Data were collected using the four FAZIA blocks arranged in a wall configuration around the beam axis, covering polar angles from about 2\textdegree~to 8\textdegree~(see Fig.~\ref{fig:GeoFazia}) and located 80~cm far from the target. 
Each block includes 16 three-layers telescopes (Silicon-Silicon-CsI(Tl)) for a total of 64 telescopes. This design allows for both $\Delta$\textit{E-E} and Pulse Shape Analysis (PSA) methods for charged particle identification. 

Each telescope covers a 2x2 cm$^2$ active area, with the thickness of the different layers being 300 $\mu$m (Si1), 300 or 500 $\mu$m (Si2), and 10 cm (CsI(Tl)), respectively. They are directly coupled to custom FEE cards, featuring the preamplifiers and the fast digital sampling stages. The whole electronics is mounted near the telescopes, as part of the block, and operates under vacuum.  
The FAZIA telescopes allow for the identification of charge and mass up to Z$\approx$25 using the $\Delta$\textit{E-E} technique~\cite{CARBONI2012251}, and up to Z$\approx$20 via Pulse Shape Analysis in the silicon detectors for fragments stopped in the first layer, with identification energy thresholds that depend on the ion charge~\cite{PASTORE201742}. 

Data were stored on disk with a global trigger condition of the hit multiplicity \textit{M}$\geq$2 to suppress the high rate of elastic events, especially in the innermost detectors. This means that single-particle events (which could be frequent due to the reduced angular coverage) are largely discarded and are only detected if they are accompanied by spurious (e.g. noise) signals exceeding the acquisition threshold. Therefore, for a consistent comparison, both experimental data and model results are shown in the following only for events with a multiplicity \textit{M}$\geq$2 of well identified particles. The overall statistics is around 5$\times 10^7$ for each system.  

\section{\label{sec:Models}Model framework}

In the following, we give a short overview of the model framework adopted in this paper to interpret and discuss the experimental results. It is a common practice at Fermi energies to assume a two-stage reaction process to describe the  production of nuclear species. In the first phase of the simulation, the projectile and the target interact leading to the formation of one or more excited primary
fragments. This represents the dynamical phase of the reaction which is extended well after the projectile-target separation allowing to take into account also the fast (dynamical) emission of particles or small clusters from non thermalized fragments. After a reasonably long time when the nuclear interaction between target and projectile is assumed to be negligible, the statistical model (afterburner phase) describing the fragment decay towards their ground state is switched on. While the statistical de-excitation is relatively well understood and generally modeled using the Hauser-Feshbach formalism~\cite{Hauser}, many approaches are still being pursued for the dynamical phase depending for example on the level of physics details which one wants to study. 
Here, our main model is the antisymmetrized molecular dynamics (AMD) model that aims at a microscopic description of nuclear reactions in terms of effective interactions and that can be thus used to constrain the nuclear equation of state (nEoS). For this reason we summarize below the essential ingredients of the AMD while we refer the reader to a more detailed presentation of the
model in ref.~\cite{Ono1,Ono2,Ono3,Ono4,Ono5}. 

AMD implements the time evolution of a multi-nucleon system which originates from two boosted nuclei in their ground state (projectile and target, with total mass number \textit{A}$_{tot}$). At each
time step, the state of the system is determined by a single Slater determinant of Gaussian wave packets for the \textit{A}$_{tot}$ nucleons~\cite{Ono1,Ono2}.  
The interaction describing the mean field in the AMD version used in this work is implemented via an effective Skyrme SLy4 force, with $K_{sat}$ = 230 MeV for the incompressibility modulus of the nuclear
matter and with $S_0=32$ MeV and $L_{sym}=46$ MeV for the symmetry energy parameters~\footnote{The symmetry energy is expanded around $\rho_0$ as $E_{sym}(\rho) = S_0 + L_{sym}\cdot\frac{(\rho-\rho_0)}{3\rho_0} + O\frac{\left[(\rho-\rho_0)^2\right]}{9\rho_0 ^2}$; a model for the symmetry energy is defined as stiff or soft depending on the value of the slope parameter $L_{sym}$.}\cite{CHABANAT1997710}, called the asy-soft parametrization. The code also includes a variant of this force corresponding to an asy-stiff nEoS with $S_0=32$ MeV and $L_{sym}=108$ MeV~\cite{IkenoPhysRevC.93.044612}.  
The latter parametrization has resulted to be slightly better in reproducing experimental data for medium-heavy mass reactions at Fermi energies~\cite{Camaiani_PhysRevC.102.044607, PiantelliPhysRevC.101.034613, PiantelliPhysRevC.99.064616,PiantelliPhysRevC.103.014603}. In the reactions here studied the sensitivity to the nEoS asy-stiffness is expected to be weak; in fact, the interacting nuclei
are autoconjugate and the possible exploration of subsaturation densities is hindered by the reduced system size.    
Therefore we opted for the original SLy4 parametrization that provides a better reproduction of the binding energies in this light nuclei region and thus gives a more reliable excitation energy as input
for the afterburner.   
In addition to the mean-field, residual interactions must also be taken into account, especially for energies higher than 10 MeV/nucleon. The residual interaction was considered in AMD using stochastic nucleon-nucleon (NN) collisions, since this effect goes beyond the mean-field approximation with a single Slater determinant. The total collision probability and the partial transition probabilities to specific states are calculated based on the differential effective cross section $(\delta \sigma/ \delta \Omega)_{NN}$ of the two-nucleon scattering. The transition probability also depends on the presence of the nuclear medium (see \cite{Lopez_PhysRevC.90.064602}). In the current version of AMD, the parametrization was adopted from~\cite{CouplandPhysRevC.84.054603}, with the screening parameter y=0.85 as in~\cite{Camaiani_PhysRevC.102.044607,PiantelliPhysRevC.101.034613,Lopez_PhysRevC.90.064602} corresponding to a slight in-medium reduction of the free NN cross section. 
\begin{figure*}[] 
   \centering 

   \includegraphics[width=2.0\columnwidth]{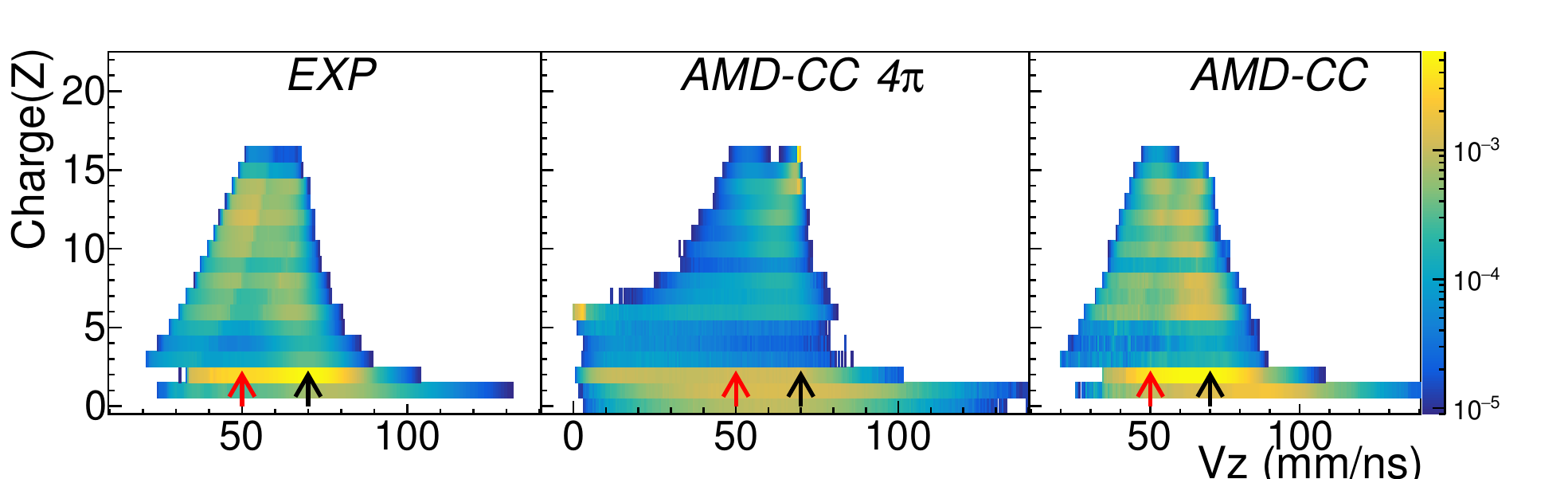}

   \caption{Experimental and model correlations, before (AMD-CC 4$\pi$) and after the filter (AMD-CC), of ion charge v.s. parallel velocity in the laboratory frame for the S$+$C reaction at 25 MeV/nucleon. The red and black arrows indicate the center of mass and projectile velocities, respectively. Counts are normalized to unity (i.e total number of events) for a better yield comparison.} 

   \label{fig:Zvz} 

\end{figure*} 
The used AMD version~\cite{Ono4, Ono5} allows for a good description of cluster emission by considering nucleon-nucleon, nucleon-cluster and inter-cluster correlations. First, each of the scattered nucleons is allowed to form a light cluster such as a deuteron, triton, $^3$He or $\alpha$ particle through collisions. The collision probability leading to a specific cluster configuration is implemented as described in Refs.\cite{IkenoPhysRevC.93.044612,OnoJPSCP.32.010076}. As a further coalescence step, light clusters (with mass number A$\leq 4$) can form bound states corresponding to light nuclei such as Li and Be isotopes (referred here as IMF) or beyond (up to around A$\approx10$). This inter-cluster correlation is introduced as a stochastic process and the residual relative momentum between clusters is set to zero if moderately separated clusters are moving away from each other with a small relative kinetic energy ($p^2_{rel}/2\mu<$ 10 MeV). When a cluster is created, the internal state of the cluster is changed to its ground state wave function. We assume the nucleons to occupy the lowest level in a harmonic oscillator potential, which is described by putting the Gaussian wave packets at the same phase-space point in AMD. 
For more details on the clusterization and inter-clusterization processes see sec.III of~\cite{PiantelliPhysRevC.99.064616} where the same implementation of the code has been used.  

For our purposes, as done in the past~\cite{Camaini_PhysRevC.103.014605, PiantelliPhysRevC.99.064616}, we verified that the dynamical phase is completed at 500 fm/c after the onset of the interaction, and the AMD calculations are stopped at this point. We have simulated about 60000 primary events for the four systems, with the impact parameter ranging from 0 to 9 fm (i.e. up to about the grazing values) according to a continuous sampling of a triangular distribution. The primary events contains, as said before, various fragments that can be excited right after the reaction (i.e.~dynamical phase).  
To simulate the statistical de-excitation, these primary fragments were treated with the Hauser-Feshbach light~(\hf) code~\cite{Baiocco054614,Baiocco2012} which then produces the final ``cold'' fragments as seen in the detectors. The Monte Carlo \hf{} code was purposely developed to study light excited nuclei. The evaporation of light charged particles (LCP, i.e. \textit{Z}$\leq$2) and
IMF (up to Be) is treated within the Hauser-Feshbach theory~\cite{Hauser} with the transmission coefficients implemented as in Ref.~\cite{Chen_PhysRevC.38.2630}.   
For each AMD primary event, 1000 secondary statistical decays were simulated with~(\hf); therefore, approximately 60 million events are available for each reaction system before the apparatus filtering phase. 
By reducing the statistical event multiplicator (from 1000 to 1), we have verified that this procedure does not introduce significant biases for the observables investigated in the following analysis.

In this paper we also report some results obtained with a second reaction model, which is the semi-phenomenological generator labelled HIPSE, successfully employed to predict various observables for
reactions at Fermi energies~\cite{MockoPhysRevC.78.024612, Vient_PhysRevC.98.044612}. Also this model follows the two-phase scenario to describe the nuclear reaction and in its dynamical part it also includes a coalescence algorithm which is relevant for the present purpose to investigate the cluster formation. We notice that the test of HIPSE with respect to cluster production leads to a limited physical information on nuclear fundamental properties.
However, we consider valuable to present some results of HIPSE for the present nuclear reactions, considering the relatively fast event production and the good overall performance showed in previous papers
\cite{MockoPhysRevC.78.024612, Dudouet2,Vient_PhysRevC.98.044612} which we refer for further details on the model.
To simulate the statistical de-excitation, the primary fragments generated by HIPSE were treated with the SIMON afterburner~\cite{DURAND1992266} which is the standard evaporation code included in the software package, also computing the Coulomb effects affecting the asymptotic trajectories of fragments. We decided to adopt the entire HIPSE+SIMON package as it is, with the reaction parameters set at the values pertinent for the measured energies and without attempting any further tuning of the evaporation decay parameters. 
Overall, we have simulated with the HIPSE package a total of 15 million (secondary) events for each system prior to the geometrical filtering phase following the same prescription for the impact parameter distribution as in AMD.

\section{\label{res}Results and Discussion}

In this section we present the results of the comparison between the model predictions and the experimental data. First we focus on global reaction  observables as a function of the projectile and beam energy. These observables will be compared with the predictions of both HIPSE and AMD in the version including clusterization which performs the best. Then we go into more details and we investigate the influence of the clustering mechanism on various observables such as the light fragment multiplicities.  
This investigation is done using the two versions of AMD, with and without clusterization that will be indicated as AMD-CC and AMD-NC, respectively. The comparison is extended also to the predictions of 
HIPSE that, as said before, implements the cluster formation (through a coalescence algorithm) during the collision.

Before the comparison, the simulated data are filtered through a software replica of the apparatus which takes into account experimental conditions and effects as for instance the geometrical efficiency, the identification thresholds and the energy resolutions. Taking as an example the reaction S$+$C at 25 MeV/nucleon, in Fig.~\ref{fig:Zvz}, the correlations between the charge
($Z$) and the longitudinal ($V_z$) velocity in the laboratory reference frame for all particles are shown for the experimental, AMD unfiltered (4$\pi$) and AMD filtered data, respectively.
By comparing the simulations in 4$\pi$ and after filtering we observe that the acceptance of the FAZIA modules does not introduce major cuts on the whole phase-space except for the low-\textit{Z}, low-Energy regions pertaining to the Carbon target region, as expected. Neutrons are not detected by the FAZIA modules, designed for charged particles.
In particular we observe that  both the QP (black arrow) and the \cm~(red arrow) regions are clearly accessible by the apparatus. This occurs also thanks to the kinematical forward focusing for reverse kinematics, even stronger at 50 MeV/nucleon. Finally, we observe that the region of quasi-elastic events focused in the peak around \textit{Z}$=$16 (in 4$\pi$), is strongly suppressed due to the \textit{M}$\geq2$ trigger condition. The small percentage of surviving \textit{Z}$=$16 events are those with a high degree of energy dissipation through particle emission from rather central events.

\subsection{\label{sec:ZandVz}Charge and velocity distributions}

\begin{figure}[!h] 
   \centering 
   \includegraphics[width=1.0\columnwidth]{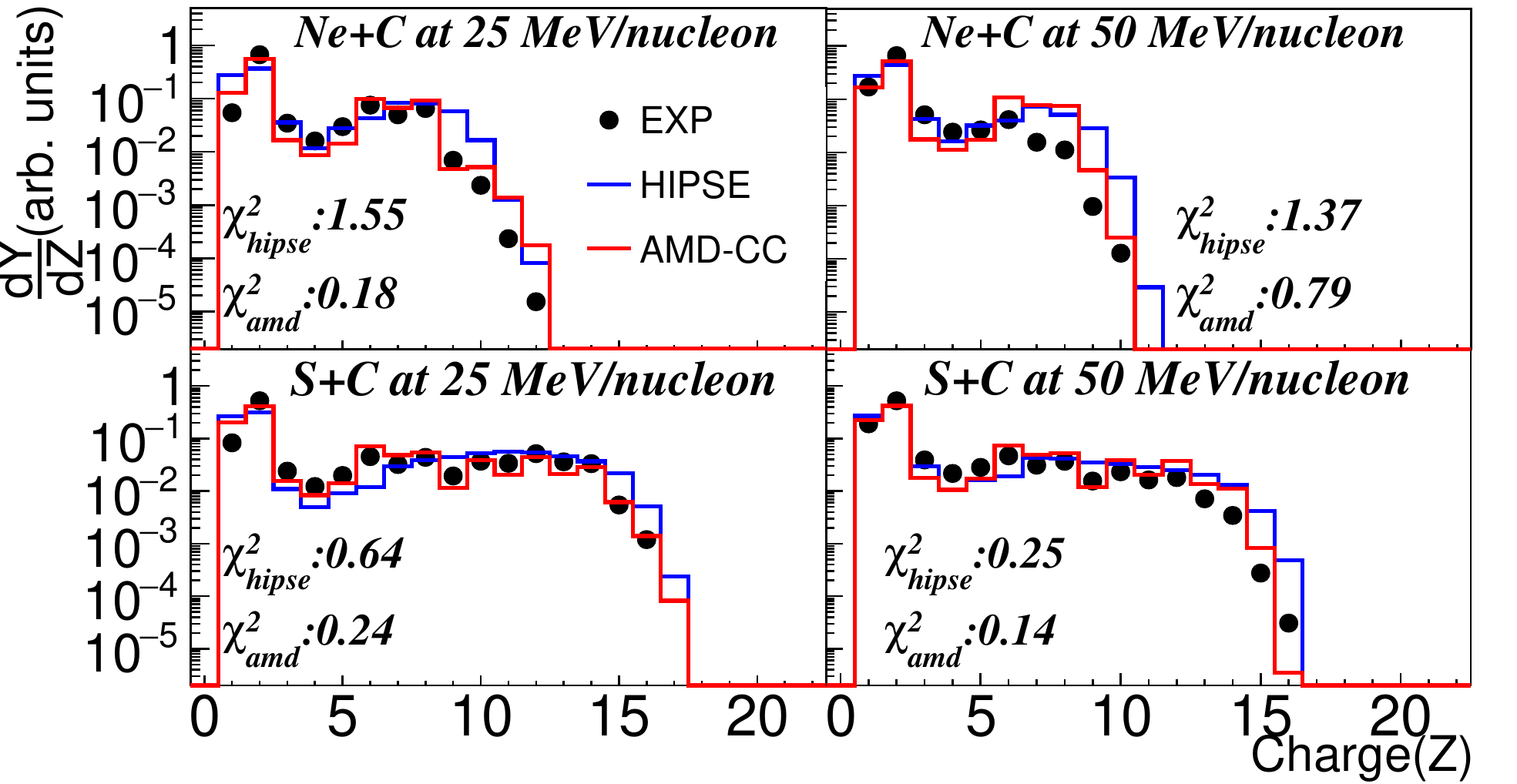}
   \caption{Charge probability distributions for the four reactions (indicated in each pad). Black points show the experimental data while continuous lines represent the AMD-CC model and HIPSE calculations, respectively. 
    } 
\label{fig:Z}  
\end{figure}

\begin{figure}[!h] 
   \centering 
   \includegraphics[width=1.0\columnwidth]{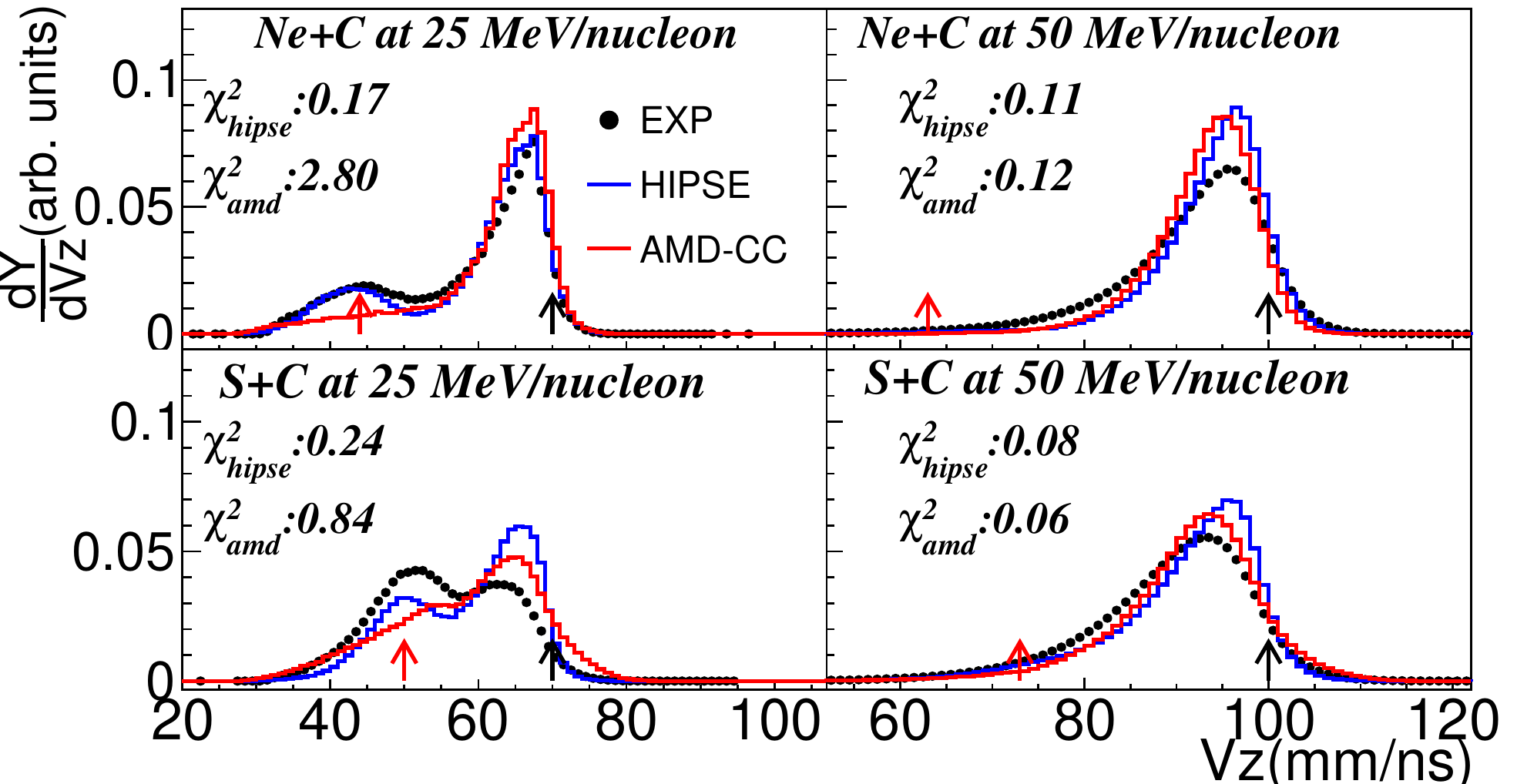}
   \caption{Velocity probability distributions of \textit{Z}$>$5 fragments along the beam axis in the lab reference frame for the four reactions (indicated in each pad).
   Black points show the experimental data while continuous lines represent the AMD-CC model and HIPSE calculations, respectively. Arrows indicate the center of mass velocity (red) and beam velocity (black).} 
   \label{fig:Vz} 
\end{figure}

In Figs.~\ref{fig:Z} and \ref{fig:Vz}, the probability distributions of the charge $Z$ and of the longitudinal $V_z$ velocity in the laboratory frame (in this case, only for \textit{Z}$>$5), are shown respectively. The model predictions (lines) are compared with the experimental data (dots) for the four measured reactions.  

Besides to the natural scaling with the projectile size, the $Z$ distributions evolve with increasing bombarding energy. At 50 MeV/nucleon the production of light species (namely protons and the IMF region) is favored possibly due to two concurrent effects: more violent dynamics at early reaction phases can produce fast fragment emissions; moreover, more dissipative collisions for mid-central events lead to more excited fragments with consequent longer secondary particle evaporation chains.
 
Both models perform rather well and reproduce the general trend of the charge and velocity distributions. However, at a closer look, differences emerge both in charge and in velocity spectra. 
In order to quantitatively test the model performance, we introduce a Pearson-like $\chi^2$ implemented as:
\begin{equation}
\chi^2=\sum_{i} \frac{(Y_i^{exp}-Y_i^{mod})^2}{Y_i^{exp}}  
\label{pearson}
\end{equation}
where $Y_i^{exp}$ and $Y_i^{mod}$ are the values of the experimental and simulated spectra in the \textit{i}th bin, respectively. The $\chi^2$ values represent the mean deviation of the
model prediction with respect to the experimental data used as the expectation value, bin per bin. The obtained $\chi^2$ of each model are reported in the panels of Fig.~\ref{fig:Z} and \ref{fig:Vz}.
Considering the charge spectra, AMD performs better than HIPSE as signaled by $\chi^2$ values always less than HIPSE. This is mainly due to the fact that HIPSE tends to produce too large fragments, in particular close to the region of the projectile charge.

AMD faithfully reproduces the charge distributions, especially at 25 MeV/nucleon; also the odd-even staggering is predicted at the correct level. The LCP yields will be discussed in detail in the following; we anticipate here that both models overpredict hydrogen isotopes while underpredict helium particles, especially at 25 MeV/nucleon. 
This abundant $\alpha$ production, not  fully described by models, is a characteristic of light N=Z systems where the underlying $\alpha$ clustering may play an important role not only in the dynamical phase but also in the following statistical decay chain~\cite{Morelli_2014, Morelli_2014_v2, CamaianiPhysRevC.97.044607,borderie2016,borderie2021}. 

The experimental $V_z$ distributions of detected fragments with \textit{Z}$>$5 (Fig.~\ref{fig:Vz}) also show a large change with bombarding energy; at 25 MeV/nucleon there are two peaks, one slightly below the projectile velocity and the other corresponding to the \cm-velocity (these relevant velocities are marked with arrows in the figure). The former peak is associated to collisions where the \textit{Z}$>$5 fragments keep memory of the projectile phase-space while the broader peak at \cm-velocity is ascribable to fusion-like processes where basically the initial momentum is fully conserved by a compound system located near the center of mass. This hot source then decays towards lighter fragments; due to the small size of the C-target, the decay of a QP and the decay of a slightly heavier fused-system lead to similar fragment size, although with different velocities. 
According to the fusion systematics~\cite{Eudes_PhysRevC.90.034609} at 25 MeV/nucleon one can expect a fraction of approximately 13\% and 4\% of fusion-like events, respectively for the S+C and Ne+C reactions.
We want to remark that these percentages are in a 4$\pi$ configuration.
These are not the ones experimentally observed in Fig.~\ref{fig:Z} and \ref{fig:Vz} strongly affected by the geometrical filter which cuts peripheral collision as shown also in Fig.~\ref{fig:Zvz}.
The depicted scenario is no longer valid at 50 MeV/nucleon where the $V_z$ distribution presents only one broad peak below the projectile velocity with a long tail extending to values around (for Neon) or even well below (for Sulfur) the \cm-velocity, signaling a smooth transition of events from semi-peripheral to semi-central collisions with a broad range of dissipated energy. The result is consistent with the expected disappearance of fusion processes according to the systematics. As signaled by the $\chi^2$ values, at 25 MeV/nucleon HIPSE performs better 
than AMD basically because AMD fails to produce the fusion-like hump around the \cm-velocity; HIPSE in particular performs very well for the Ne-induced reaction.
Both models well describe the disappearing of fusion-like events at 50 MeV/nucleon and globally reproduce the velocity spectrum with similar success (similar $\chi^2$).

By summarizing this part, we comment that both models have specific points of strength. AMD describes better than HIPSE the charge distributions but fails in populating the velocity regions close to the CM perhaps due to the underestimation of the cross section of fusion-like reactions, better modeled by HIPSE. We underline that also within AMD there would be the possibility to somehow play with parameters (such as the NN cross section or the wave-packet widths) in order to improve the agreement with the experimental observables. This tuning, however, is beyond the scope of this work and has not been attempted; as mentioned, in this work AMD has been run with the standard parameters as already done for other heavier systems~\cite{PiantelliPhysRevC.101.034613,Camaini_PhysRevC.103.014605}.

\begin{table*}[]
\caption{Event sorting based on the heavy fragment multiplicity (\textit{M}$_{HF}$) for the two reactions. The experimental data (EXP) are shown together with both model predictions (HIPSE and AMD). The shown percentages are referred to the total number of events. Errors are the sum of the statistical component and of the identification procedure (less than 1\% for each presented value).}
\vspace*{0.2cm}
\centering
\label{table:Mbig1}
\renewcommand{\arraystretch}{2.0}
\setlength{\tabcolsep}{8pt} 
\begin{tabular}{cl|ccc|ccc|ccc|}
\multicolumn{2}{c|}{\multirow{2}{*}{}} & \multicolumn{3}{c|}{M$_{HF}=0$ (\%)} & \multicolumn{3}{c|}{M$_{HF}=1$ (\%) } & \multicolumn{3}{c|}{M$_{HF}\geq2$ (\%) }        \\ \cline{3-11} 
\multicolumn{2}{c|}{}                  & EXP        & HIPSE  
& AMD-CC         & EXP        & HIPSE        & AMD-CC        & EXP           & HIPSE         & AMD-CC           \\ \hline
\multirow{2}{*}{25 MeV/nucleon}    & Ne+C    & 58         & 45          & 45        & 42         & 55          & 55        & \textless{}1 & \textless{}1 & \textless{}1 \\
                             & S+C     & 33         & 20          & 31        & 62         & 79          & 63        & 5            & 1            & 6            \\ \hline
\multirow{2}{*}{50 MeV/nucleon}    & Ne+C    & 84         & 62          & 44        & 16         & 38          & 56        & \textless{}1 & \textless{}1 & \textless{}1 \\
                             & S+C     & 54         & 44          & 30        & 44         & 55          & 66        & 2            & \textless{}1 & 4            \\ \hline
\end{tabular}
\end{table*}
\subsection{\label{sec:Multiplicities} Fragment Multiplicity and Clustering}

A global classification of the events can be performed also on the basis of the heavy fragment (HF, defined as \textit{Z}$>$5) multiplicity. The event sharing according to the HF multiplicity is reported in table~\ref{table:Mbig1}, for both experimental data and model predictions. 

In the experimental data, we observe a significant amount of \textit{M}$_{HF} =$0 events, increasing with the bombarding energy and more abundant in the light Ne+C system. Various effects contribute to this result. The trend with bombarding energy can be affected by the increasing role of multifragmentation channels which reduce the final production of ions equal or heavier than Carbon. Also, Carbon ions (corresponding to arbitrary limit for the adjective ``heavy") are surviving with larger probability in the heavier S+C system than in Ne+C collisions. 

Last but not least, the limited acceptance which slightly differs for the two systems can have a role. In a complementary way, the cases with one HF increase with the system size while they decrease with beam energy. The events with \textit{M}$_{HF}\geq$2 are minor for these systems also due to the limited detector acceptance that plays an increasing role for manifold events. However, this category of events strongly increases with the system size; indeed, any kind of break up from S-like sources tends to produce two or more fragments with \textit{Z}$>$5 more frequently than in the case of Ne+C source. 

On the model side, the partitioning is only partially reproduced. Overall, there is an overproduction of \textit{M}$_{HF} =$1 events in all reactions (25 and 50 MeV/nucleon) and the correlated underestimation of zero-HF events for both HIPSE and AMD. The only exception is represented by the S+C system at 25 MeV/nucleon where AMD is very close to the experimental values (absolute differences within 2\% for all classes of events). For Ne+C at 25 MeV/nucleon, on the other hand, the HIPSE and AMD values are very similar and differ from the experimental values by 13\% for both \textit{M}$_{HF}=$0 and 1 in absolute terms. At 50 MeV/nucleon (see bottom rows of table~\ref{table:Mbig1}), the picture changes and the differences between HIPSE and AMD are significant. 
Anyway, the inspection of table~\ref{table:Mbig1} suggests that both models struggle to reproduce the experimental data: too many events with one large residue, that survive the efficiency cuts, are produced. 
The result which mainly strikes is that the percentages of M$_{HF}$=0 and M$_{HF}$=1 for the AMD-CC calculations are virtually independent of the bombarding energy for the two systems, whereas HIPSE, though not reproducing the absolute values, follows the trend of the experimental data. This could be partially the consequence of the disappearing of the fusion-like yield in HIPSE from 25 to 50 MeV/nucleon that is not the case in AMD. This disappearance provokes a big change in the 4$\pi$ yield of heavy fragments and so in the HF classes; in AMD the fusion-like seems to be always less and their suppression with beam energy is therefore negligibly affecting the event classes.  We are aware that the sharing of “heavy” fragment (larger than Carbon) multiplicity is ruled not only by the details in the fragment formation and deexcitation but also by the reliable implementation of the apparatus filter. In particular the class M$_{HF}=$0 (and in turn M$_{HF}=$1) is strongly impacted by the filter; indeed this class is weakly populated for Ne$+$C and almost empty for S$+$C in the 4$\pi$ calculations for both models. 

\begin{figure*}[] 
   \centering 

   \includegraphics[width=2.0\columnwidth]{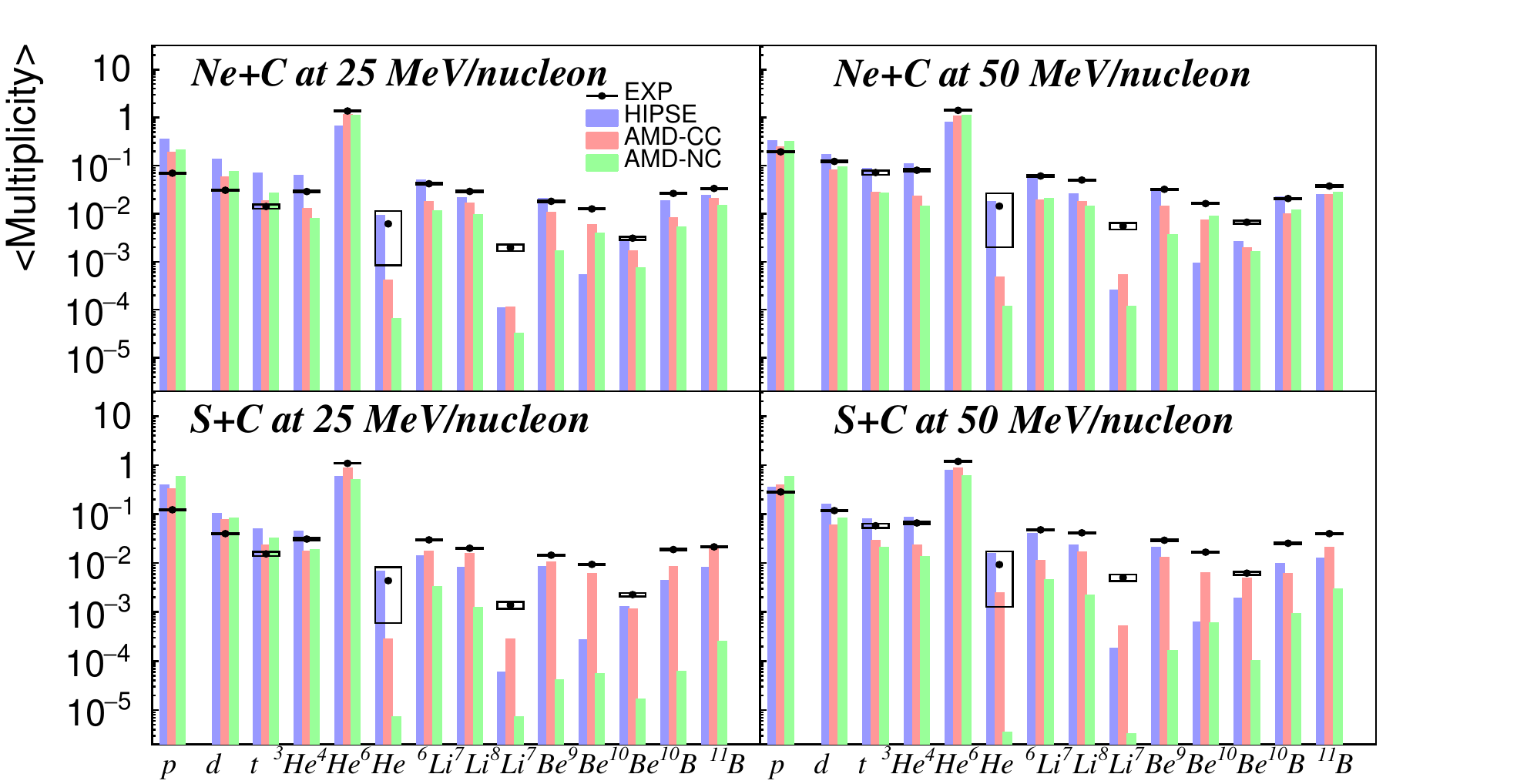}

   \caption{Proton and cluster multiplicities (\textit{Z}$<$6), after  the secondary decay, in \neo{}+\ca{} (upper panels) and \solf{}+\ca{} (bottom panels) reactions. Black points represent the experimental data while red bars represent AMD+HFl with cluster formation, green bars are for AMD+Hfl without cluster formation and blue bars represent HIPSE+SIMON simulations. The horizontal bars attached to the points are drawn to guide the eye as an indicator with respect to the models quotation. The y-errors are within the point dimension or are shown as rectangular shapes surrounding the points. Errors are the sum of the statistical contribution and of the identification procedure. } 

   \label{fig:Clusters} 

\end{figure*} 
\begin{figure*}[] 
   \centering 

   \includegraphics[width=2.0\columnwidth]{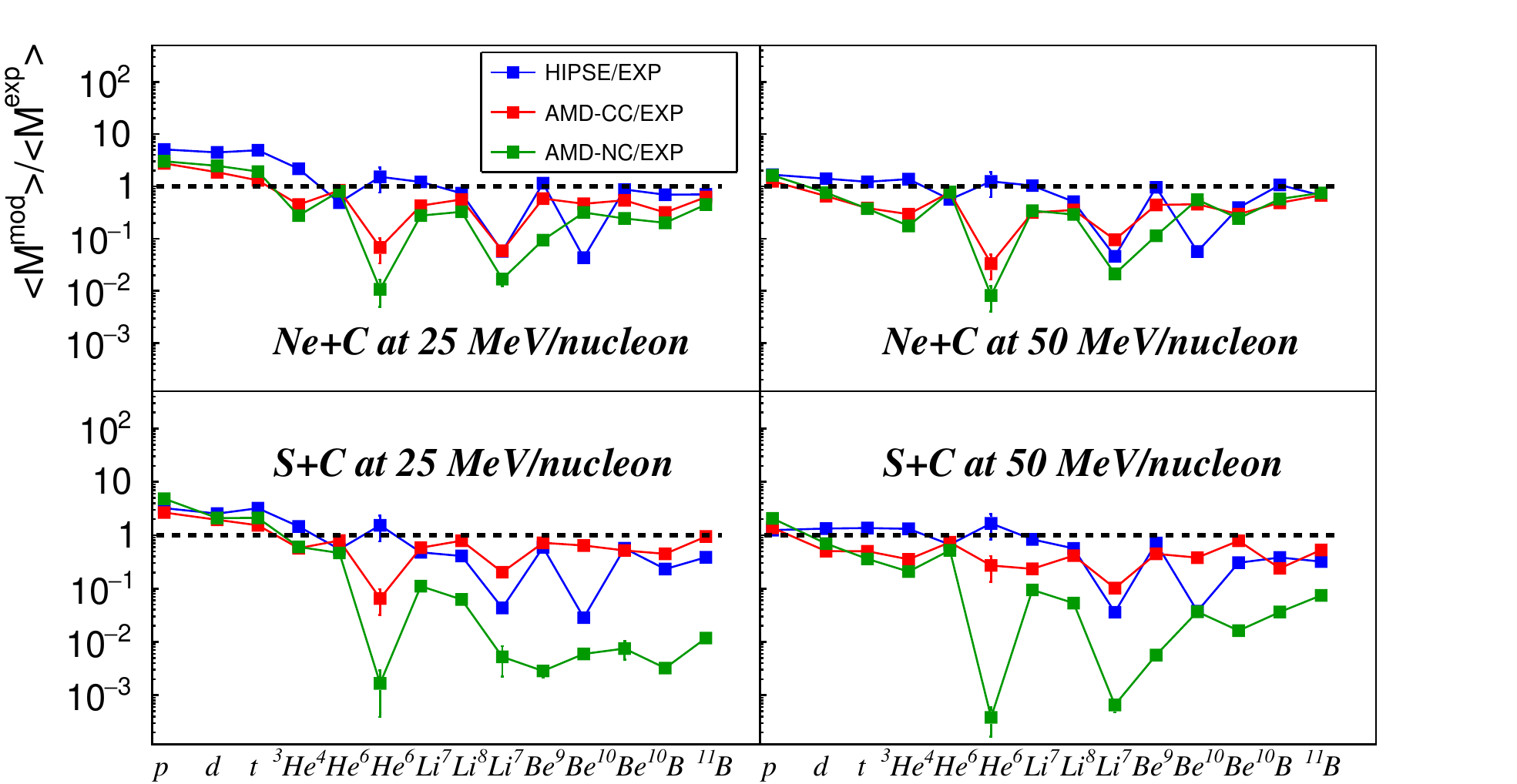}

   \caption{Ratio of model-predicted and experimental multiplicities for \textit{Z}$<$6 products in \neo{}+\ca{} (upper panels) and \solf{}+\ca{} (bottom panels) reactions. Simulated data refer to the end of the evaporation phase. Errors are indicated when exceeding the point size and lines are drawn to guide the eye. The black dashed line indicates full agreement between data and model predictions.} 

   \label{fig:Clusters3} 
\end{figure*} 

A more detailed overview of the reactions can be obtained also by comparing the average LCP and IMF multiplicities and their ratio between the experimental data and model prediction, as shown in
Fig.~\ref{fig:Clusters} and Fig.~\ref{fig:Clusters3}. Figure~\ref{fig:Clusters} shows the absolute values of multiplicities for the investigated reactions while Fig.~\ref{fig:Clusters3} is built forming the ratios of experimental and model-predicted multiplicities. In the latter case, a black dashed line is drawn to indicate full agreement between data and model predictions. 
Data show the dominant emission of $\alpha$ particles, expected in these light \textit{N}=\textit{Z} systems: on average, more than one $\alpha$-particle is found in each collected event.  
We observe that Hydrogen particles and $^3$He are a factor 2-3 more produced for the higher energy reactions; a larger production of fragments at 50 MeV/nucleon is seen also for Li and Be isotopes. 

If we consider the models we see that the very general pattern is caught by both of them with also some details (e.g. the larger yields at 50 MeV/nucleon, the production of Li and Be larger in
Ne+C than in S+C) being well reproduced. However, in absolute terms, for both models it must be noted the overproduction of Hydrogen isotopes (in particular protons at low beam energy) and the underestimation of $\alpha$ emission, as already mentioned with respect to Fig.\ref{fig:Z}, which reaches about 50\% for HIPSE for the 25~MeV/nucleon collisions.
For the IMFs, we note that both models tend to underpredict their production in all cases especially at the higher bombarding energy. A remarkable case is $^8$Li, which is a factor 10 underpredicted by AMD and HIPSE. In general, however, AMD performs better than HIPSE apart from some specific cases ($^6$He).

Now we investigate the effect of the clusterization process in AMD by running the simulation turning off (AMD-NC) the cluster option. 
The output of this calculation is also reported in Fig.\ref{fig:Clusters} and Fig.\ref{fig:Clusters3}. We can clearly notice how the cluster option is in general important for the heavier IMF
production (see red and green lines in Fig.\ref{fig:Clusters3}) but it is crucial especially for the heavier system: indeed, in S+C the inclusion of the cluster effects in AMD
can increase the IMF yields up to a factor of around 100 for the S+C reaction at 25 MeV/nucleon, much improving the performance of the simulation. 
The system mass dependence of clustering between the \neo{} and \solf{} beams is clearly explained on the basis of total available mass. In the second case 12 more nucleons are
involved, increasing the NN collision probability and therefore the cluster formation. Also, as expected, the proton multiplicity decreases significantly by turning on the clusterization process in every system, which is naturally understood because nucleons are consumed to form clusters. 

\subsection{\label{sec:AngandEnergy1}Energy and angular distributions of LCPs}
\begin{figure*}[] 
   \centering 
   
   \includegraphics[width=2.0\columnwidth]{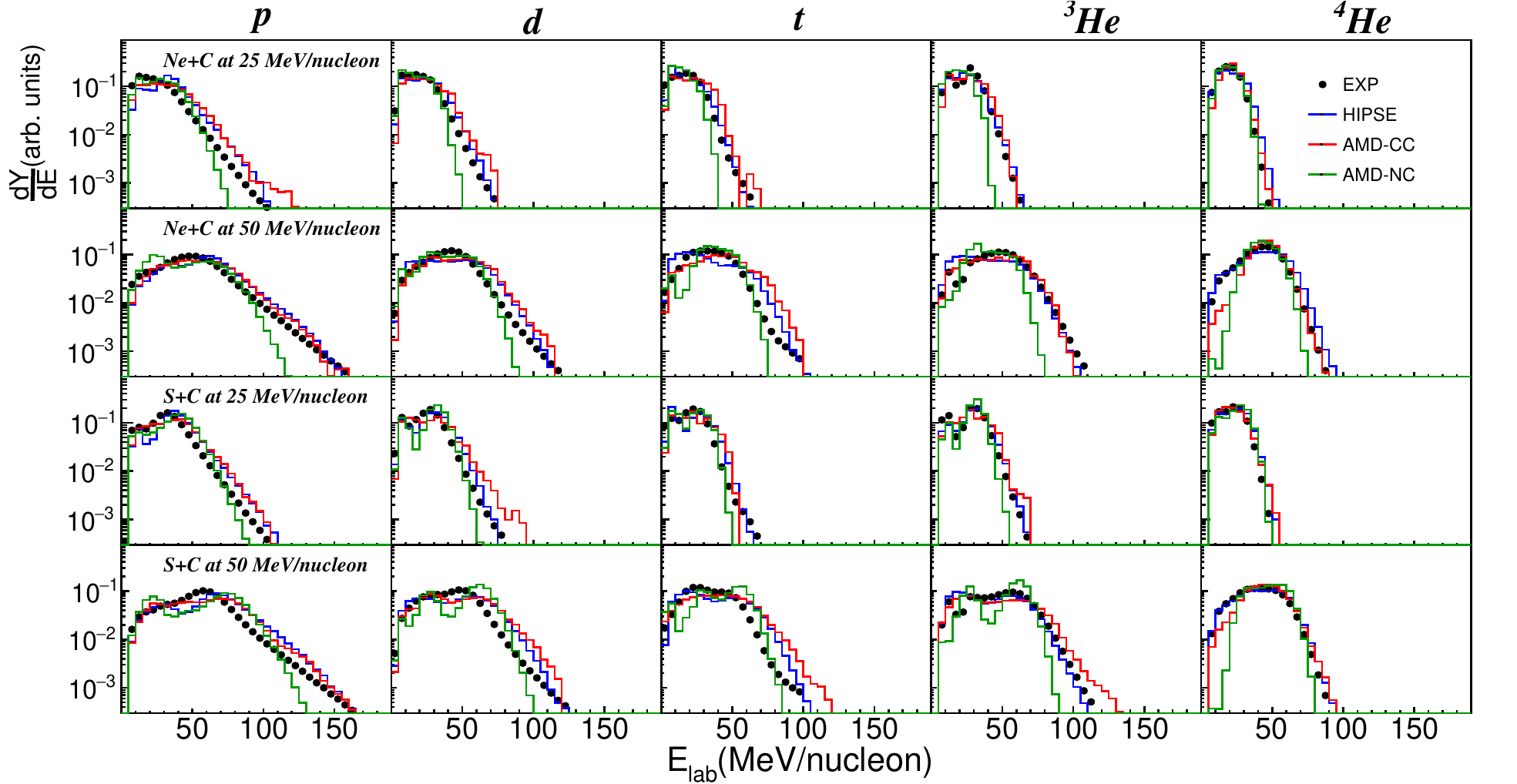}
  
   \caption{ LCP (\textit{Z}$\leq$2) energy distribution for \neo{}+\ca{} and \solf{}+\ca{}. Black points represent the experimental data while the blue, red and green lines represent
HIPSE, AMD+HFl with cluster and AMD+HFl without cluster  calculations, respectively. Spectra are normalized to the integral for a better shape comparison.} 

   \label{fig:ELCP} 

\end{figure*}
\begin{figure*}[] 
   \centering 
   
   \includegraphics[width=2.0\columnwidth]{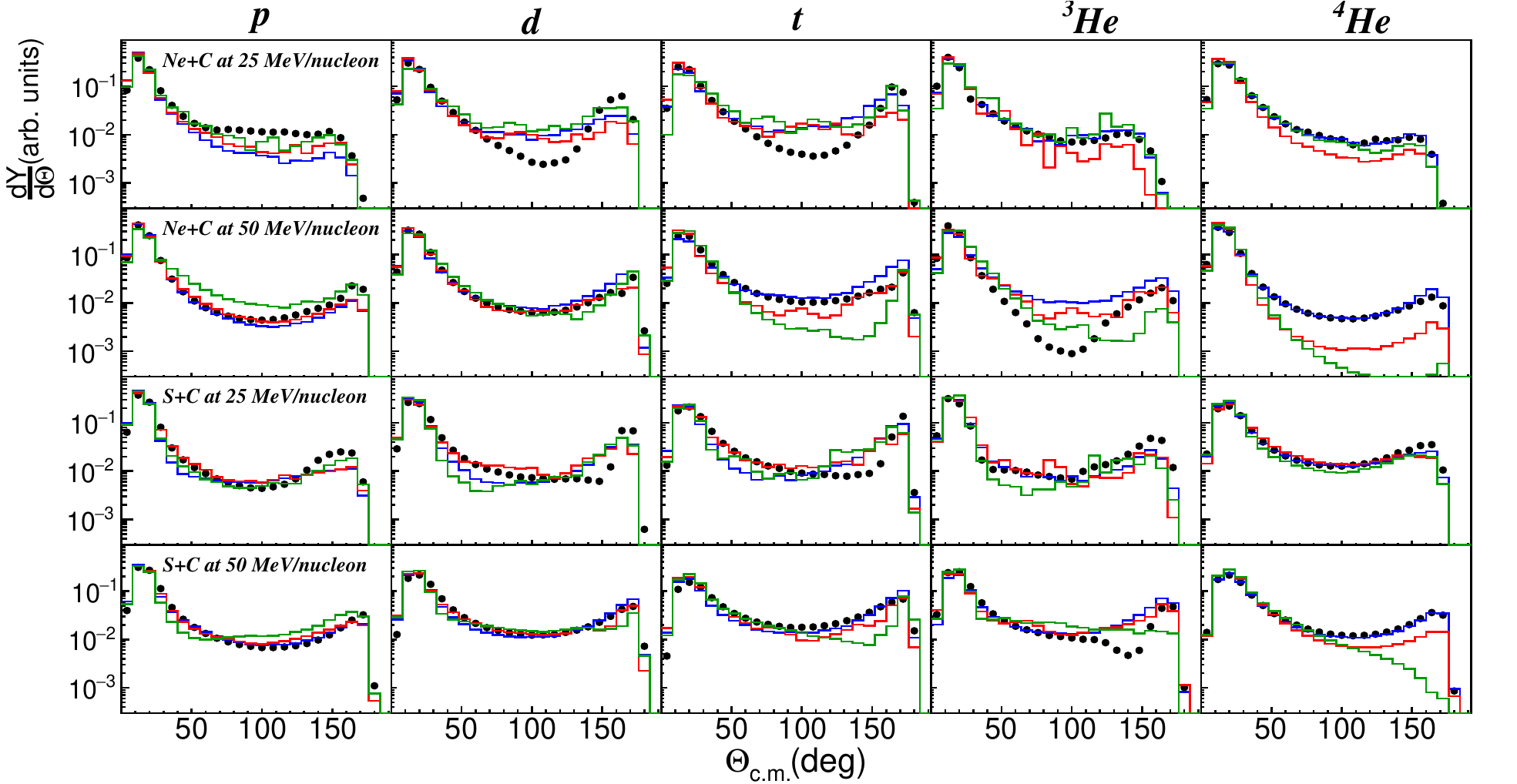}
   
   \caption{ LCP (\textit{Z}$\leq$2) angular distribution in the c.m. reference frame for \neo{}+\ca{} and \solf{}+\ca{}. Black points represent the experimental data while the blue, red and green lines represent HIPSE, AMD+HFl with cluster and AMD+HFl without cluster calculation, respectively. Spectra are normalized to the integral for a better shape comparison.} 

   \label{fig:THLCP} 

\end{figure*} 
In Fig.~\ref{fig:ELCP} the comparison between experimental data and model predictions for the energy distribution of LCPs in the laboratory frame is shown. In this figure, columns represent different
selected isotopes while rows refer to a specific system and energy as indicated in the leftmost (proton) panels.  The same applies also to the models (HIPSE and AMD with cluster cases) which overall reproduce quite well the shape of the energy distributions for all the LCPs. The AMD model was also run without the cluster correlation turned on to highlight the influence of this latter mechanism on the shape of the energy and angular distributions.

In Fig.~\ref{fig:THLCP} the comparison between experimental data and model predictions for the center of mass angular distribution are instead shown for the same LCPs. The angles are plotted in the \cm{}
frame to better highlight the different contributions which would be otherwise compressed in the laboratory frame due to the limited angular coverage of the apparatus. The labels and systems are organized as in Fig.~\ref{fig:ELCP}. 

The first observation on Figs.~\ref{fig:ELCP},~\ref{fig:THLCP} is the strong improvement obtained for the phase-space populations of LCP when including in AMD the cluster correlations. This result is clear
for both kinetic energies and angular distributions. In general, the inclusion of clustering produces a broadening of the particle phase-space which otherwise results to be too narrow and confined to the possible original main sources (a two-component structure is present in the energy spectra in AMD without the cluster option). 
As for the energy distributions, a strong effect appears at the higher beam  energy where the long tails can be reproduced for all LCP only by including clustering. Somewhat related to the broadening in energy, we also appreciate the sizable improvement of the angular distributions (particularly evident in the proton and $\alpha$ cases): the emissions at large angles are not reproduced without explicitly incorporating the clusterization process. 
\begin{figure*}[!t]  
   \centering 
   
   \includegraphics[width=2.0\columnwidth]{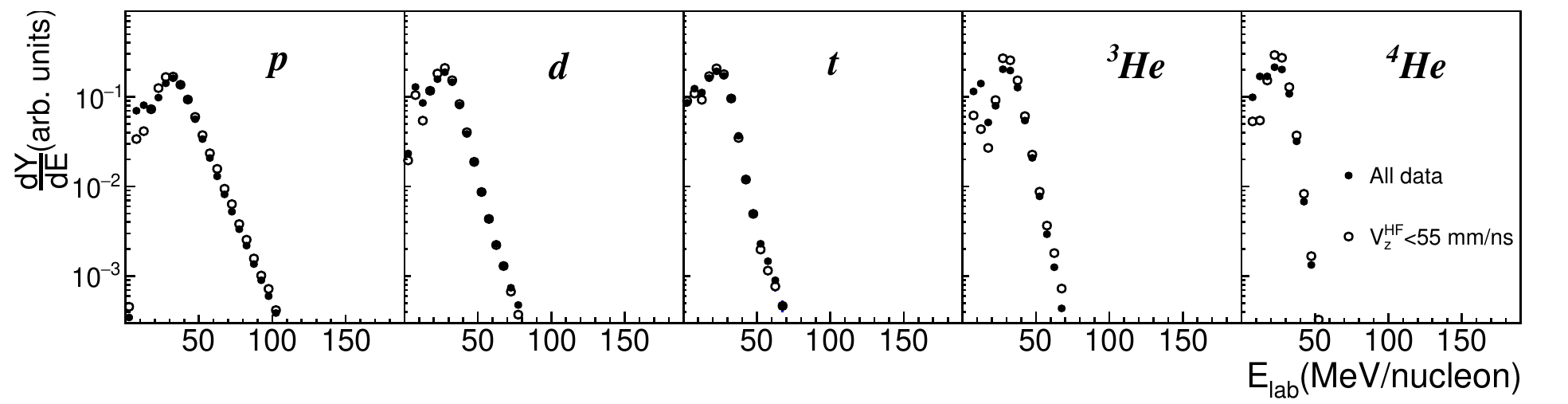}
  
   \caption{LCP (\textit{Z}$\leq$2) energy distribution for \solf{}+\ca{} at 25 MeV/nucleon. Solid points represent the experimental data of Fig~\ref{fig:ELCP} while empty pointy represents the same data after a big fragment velocity cut (i.e. V$_z <$ 55 mm/ns). } 

   \label{fig:SC25_exp_vs_exp} 

\end{figure*} 
In principle, one can argue that the scarce reproduction of the fusion-like events by
AMD (Fig.~\ref{fig:Vz}) could somehow affect the previous observation. To check this point we
applied a cut on the heavy fragment velocity to select events more likely corresponding
to incomplete fusion events; no sizable changes of the LCP spectra are found, especially
on the high energy tails, with respect to the complete dataset. This can be observed in
Fig.~\ref{fig:SC25_exp_vs_exp} where we present, only for one system, the LCP energy spectra detected in coincidence with HFs with a velocity cut V$_z <$55 mm/ns (empty points) and the full data (solid points, as in Fig.~\ref{fig:ELCP}) without any condition. Our current understanding is again based on the close similarity of the phase-space for very dissipative transfer reactions and fusion like events. Perhaps the AMD model failure is more related to the yield of fusion like events than on the features of the corresponding observables.
\begin{figure}[]  
   \centering 
   
   \includegraphics[width=1.0\columnwidth]{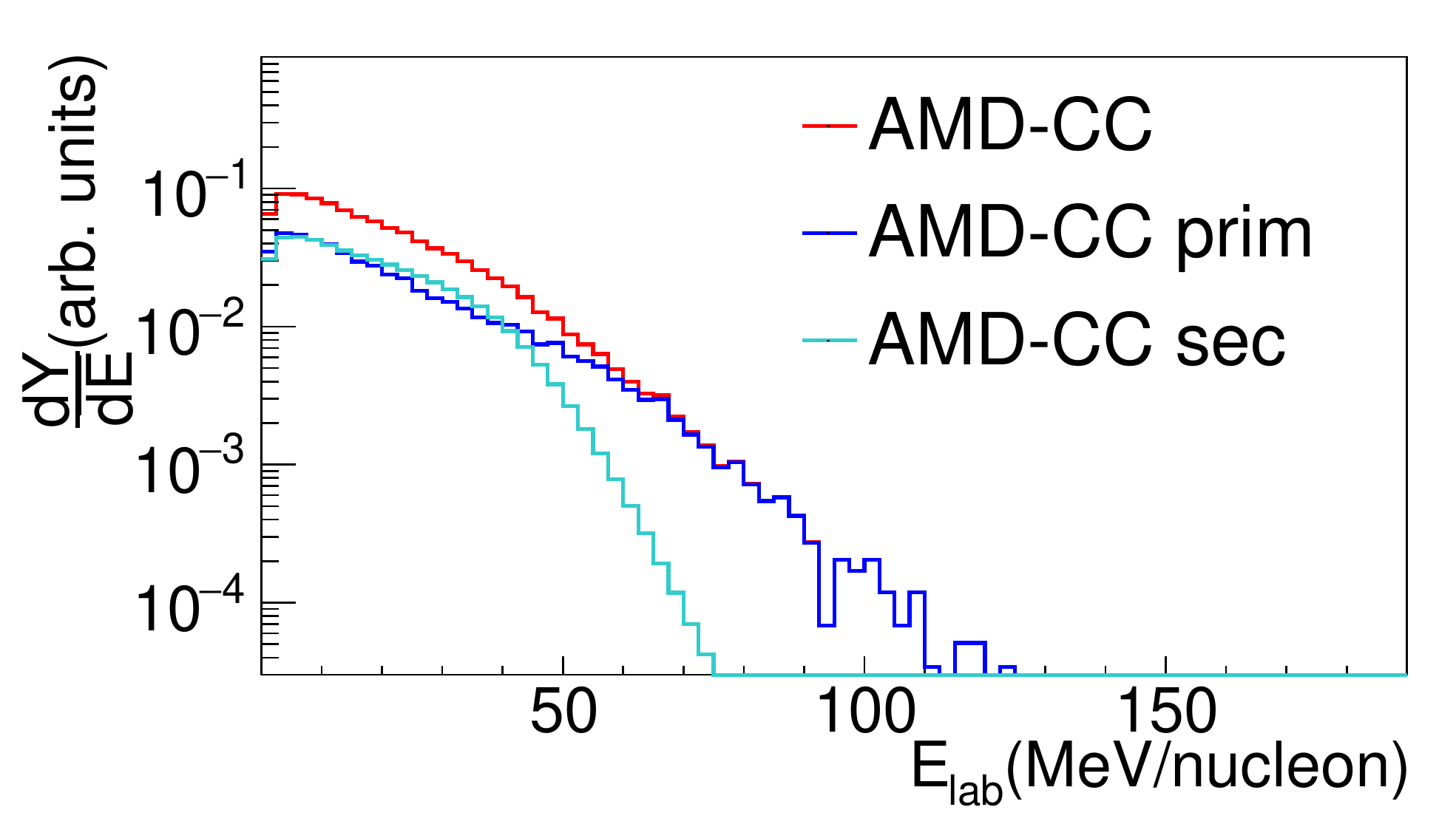}
  
   \caption{Model calculations. Proton energy distribution for  \neo{}+\ca{} at 25 MeV/nucleon in $4\pi$. The lines represent various components of the AMD-CC simulation. The red, dark blue and light blue are the total (AMD+\hf{}), AMD only (primary) and \hf{} only (secondary) components, respectively. Primary and secondary spectra are normalized to the entries of the total events of AMD+\hf.} 

   \label{fig:ELCP_prim_sec} 

\end{figure} 

\begin{figure}[!b] 
   \centering 
   
   \includegraphics[width=1\columnwidth]{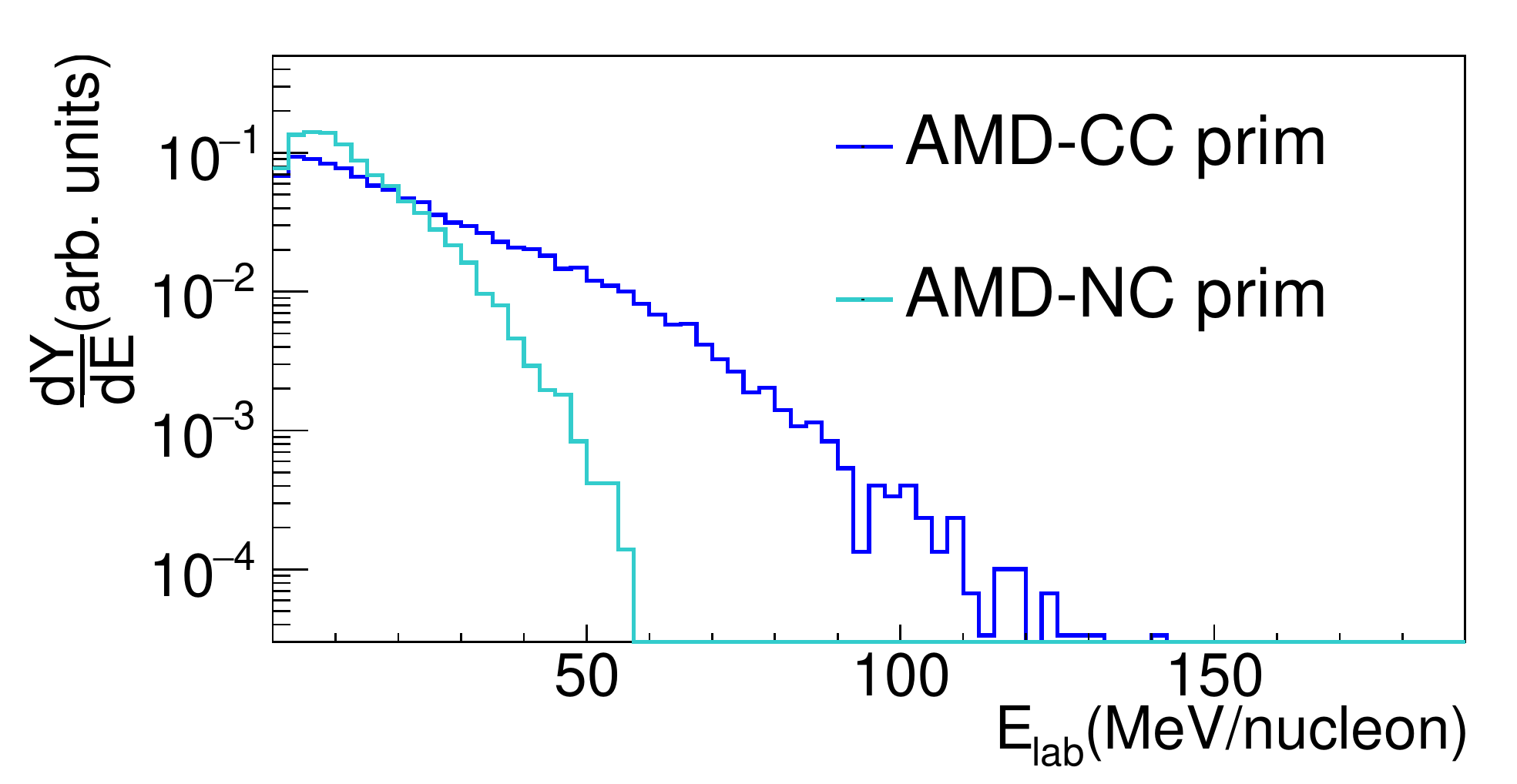}
   \includegraphics[width=1\columnwidth]{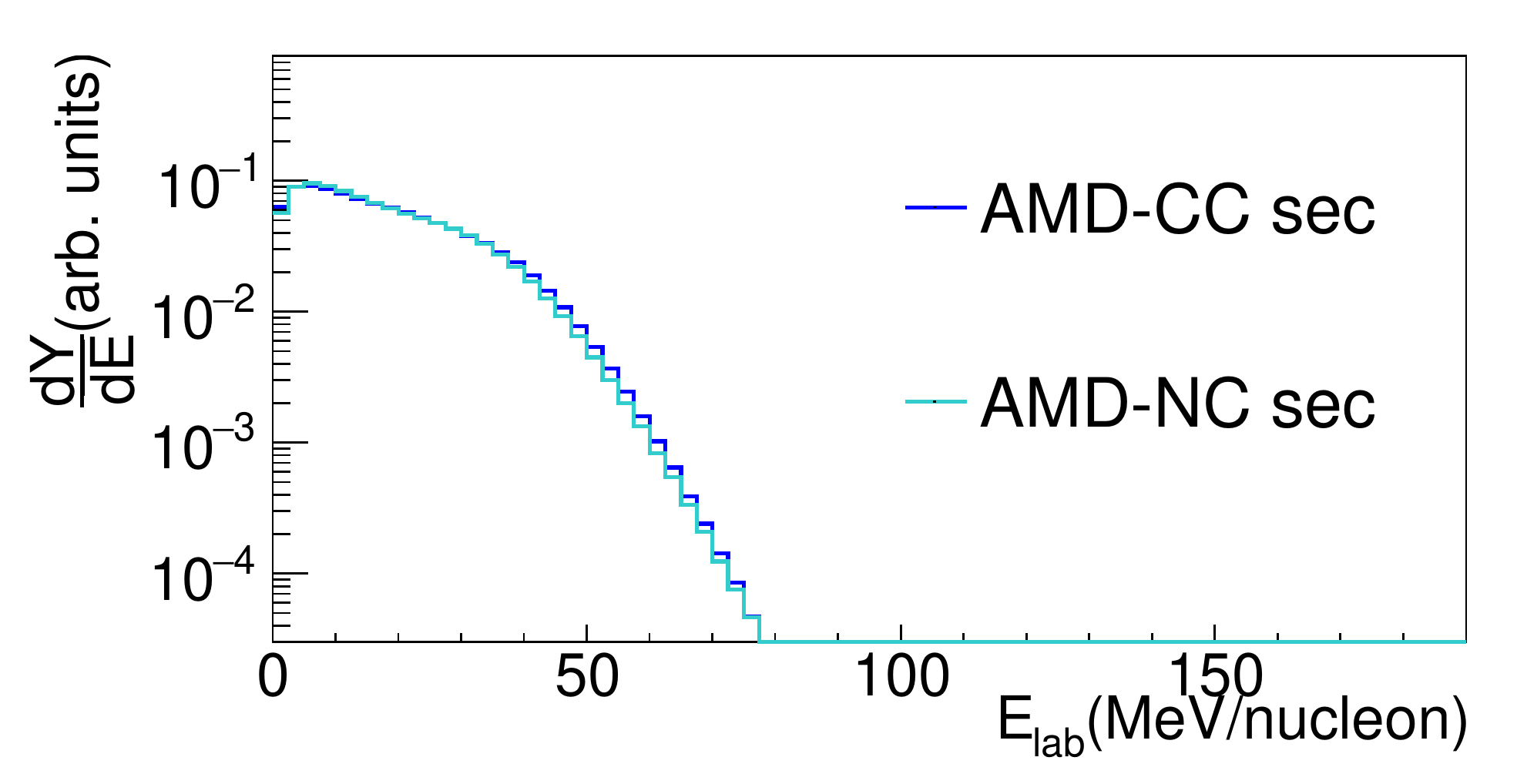}
   \caption{Model calculations. Proton energy distribution for \neo{}+\ca{} at 25 MeV/nucleon in $4\pi$.
   On top the primary component (AMD only ), with cluster and no cluster, is represented while on bottom the secondary (\hf \:only) is shown. 
 Spectra are normalized to the integral.}

   \label{fig:ELCP_prim_cc_vs_nc} 

\end{figure}

Therefore, while a direct impact on the phase-space of fragments (e.g. Helium isotopes) is obviously expected when including clustering in AMD, it might seem counterintuitive to see large effects also on single nucleons as protons for which, in particular, we observed a strong influence of clustering on their velocity/energy and angular distributions.  We checked that the high energy tail in AMD-CC derives from the primary/dynamical protons sources and is not an indirect effect of the evaporation. This can be seen for example for \neo{}+\ca{} at 25 MeV/nucleon in Fig.~\ref{fig:ELCP_prim_sec} where we plotted the primary and secondary components extracted from AMD and \hf{} in the $4\pi$ configuration (no filtering) to take advantage of the greater statistics. Evidently, the energetic protons come from the ``dynamical'' phase of the simulation. To better investigate this point, in Fig.~\ref{fig:ELCP_prim_cc_vs_nc}, we also present for the same reaction the primary (upper panel) and secondary (bottom panel) proton spectra when using AMD with (dark blue line) and without clustering (light blue line). It is clear from this figure that the significant change of the proton energy is related to the different dynamics when the cluster formation is included and not to the further evaporative decay which is basically unaffected. To explain this effect, details of energy and momentum conservation must
be taken into account as follows. 
Without the cluster option, too many single protons are produced (see Fig.~\ref{fig:Clusters}) that are emitted with smaller average kinetic energies, as suggested and argued by Refs.~\cite{ONO2019139,REISDORF2010366}. When clusterization is present, the energy released by forming a cluster is generally shared between the cluster itself and the other
collision partner, such as for example a proton. Due to the mass difference, the energy and momentum conservation results in a large kinetic energy of the proton. Energy and momentum conservation is achieved by adjusting only the relative momentum between the two particles (e.g. a proton and an $\alpha$-cluster), so other nearby particles are not modified.
Thus, the kinetic energy of the light species with clusterization can reach (indirectly) larger values than in the elementary two-nucleon collisions without clusterization. 
\begin{figure*}[] 
   \centering 
   
   \includegraphics[width=2.0\columnwidth]{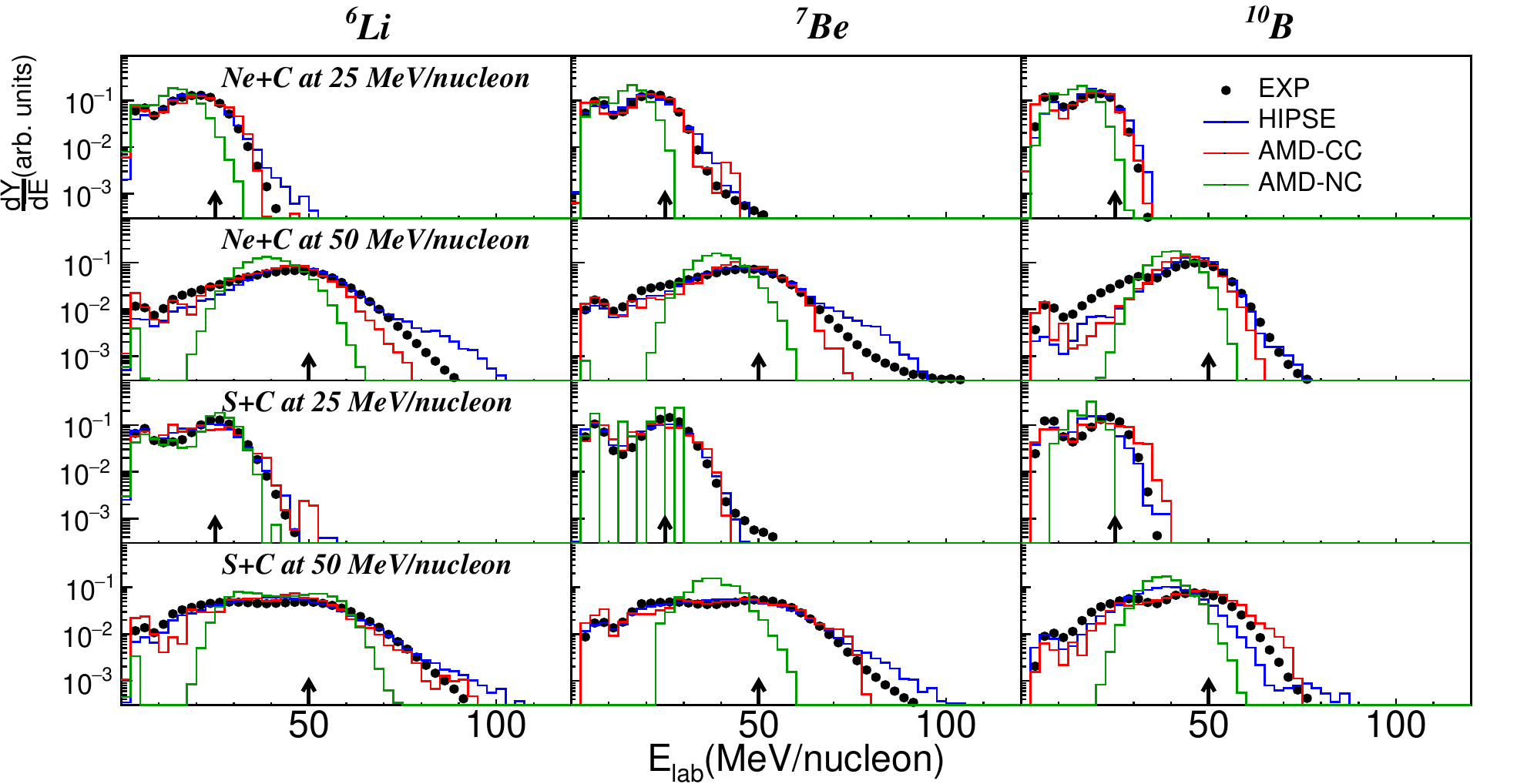}

   \caption{ IMF (2$<$\textit{Z}$\leq$5) energy per nucleon spectra for \neo{}+\ca{} and \solf{}+\ca{}. Black points represent the experimental data. Model predictions, at the end of the evaporation phase, are reported for HIPSE (blue line), AMD+\hf{} with clusterization (red lines) and AMD+\hf{} without clusterization (green lines). The black arrows indicate the beam energy position in each panel. Spectra are normalized to the integral for a better shape comparison.}

   \label{fig:EIMF} 

\end{figure*} 

\subsection{\label{sec:AngandEnergy2}Energy and angular distributions of IMFs}

\begin{figure*}[] 
   \centering 
   
   \includegraphics[width=2.0\columnwidth]{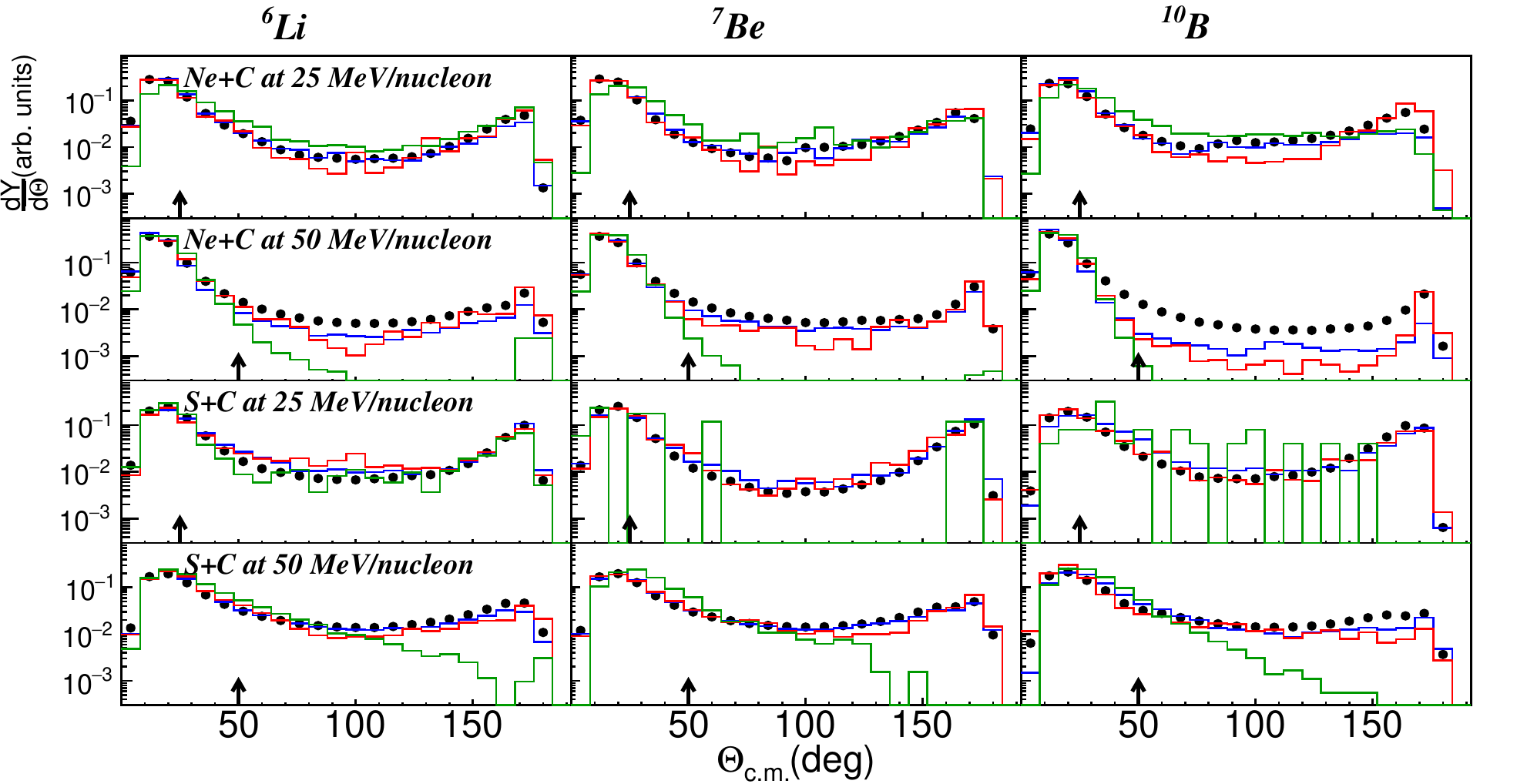}
 
   \caption{
   IMF (2$<$\textit{Z}$\leq$5) angular distribution in the c.m. frame for \neo{}+\ca{} and \solf{}+\ca{}. Black points represent the experimental data. Model predictions, at the end of the evaporation phase, are reported for HIPSE (blue line), AMD+\hf{} with clusterization (red lines) and AMD+\hf{} without clusterization (green lines). Spectra are normalized to the integral for a better shape comparison.} 

   \label{fig:THIMF} 

\end{figure*}

The analysis performed for the light particles can be extended to the IMFs. In Figs.~\ref{fig:EIMF} and~\ref{fig:THIMF} we show the laboratory energy per nucleon spectra and the \cm{} angular
distributions of selected Li, Be and B isotopes. In the collisions at 25 MeV/nucleon it is rather evident that these IMF are produced from a projectile-like source: the distributions have a main peak around the projectile velocity (indicated in all panels with an arrow); a smaller peak is also present at lower energies corresponding to fragments produced from target-like and/or \cm~sources.  
Instead, at 50 MeV/nucleon, one observes a significant broadening of the distributions for both Ne and S induced reactions, signaling the contribution of events with very different energy dissipation, from
peripheral (located next to the projectile velocity) to more central ones featuring low momenta.  

The comparison with the models again underlines the fundamental role played by the cluster correlations in reproducing the correct energy spectra. Basically, for all species and reactions the inclusion of
cluster in AMD strongly improves the agreement with the experimental data, especially at high energy where dynamical effects (i.e. related to the interaction phase) are more important.   
In the version without cluster correlations almost all fragments are produced in the projectile region with little or no yield coming from around the \cm. 
The improvement is less important at the lower energy of 25 MeV/nucleon for both projectiles. 
Finally we observe that also HIPSE performs well, although at 50 MeV/nucleon it predicts an energetic component for Li and Be not present in the experimental data. Indeed, as explained before, HIPSE explicitly includes NN collisions which have been tuned on various experimental data at Fermi energies.

Also the comparison of the angular distributions reported in Fig.~\ref{fig:THIMF} reinforces the evidence for the need of cluster correlations when modeling nuclear reaction dynamics. The improvement of AMD-CC is moderate at 25 MeV/nucleon where it can be better appreciated for the heaviest IMF; instead, the effect is strong
at 50 MeV/nucleon where the yields at backward \cm-angles are completely failed without the cluster contribution, for all fragments.    

A detailed inspection of the AMD outputs with and without clustering shows that the improvement of the IMF production has two origins. One is direct and it is the increased production rate of the primary IMFs, with a better phase space distribution; the other is the enhanced population of these fragments at lower excitation energies which thus
favors the survival probability of the excited IMFs by reducing their secondary decay as observed also in~\cite{TIAN_PhysRevC.95.044613,Tian2_PhysRevC.97.034610}.

\section{\label{sec:concl}Summary and Conclusions} 

Four asymmetric reactions, $^{32}$S+$^{12}$C and $^{20}$Ne+$^{12}$C at 25 and 50 MeV/nucleon, were measured with a well performing array of telescopes allowing complete isotopic identification of most of the produced species. Global characteristics of the reactions, that evolve with both system mass and beam energy have been studied. 
We focused in particular on LCP and IMF observables, in order to discuss the role of clustering during the interaction phase by means of comparison between model predictions and  experimental results. To this purpose we used two models describing both the interaction phase that produces the primary fragments and the statistical evaporation that leads to the final distributions to be compared with the experimental ones. The models are AMD+\hf{} and HIPSE+SIMON, suitable for predicting the behavior of heavy-ion reactions at Fermi energies. In particular, to our knowledge for the first time, data for light systems different from C+C were compared with the current AMD model, acknowledged as one of the most performing and well physically based reaction codes for the present energy domain. AMD has been run in two versions, with and without cluster formation, to highlight the role of the clustering in the measured reactions. 
As a general comment, the charge and velocity distributions of all measured reaction products are rather well described by both models and the evolution of these observables with the system mass and beam energy is nicely reproduced. A closer inspection reveals some discrepancies: AMD tends to underpredict fusion-like events while HIPSE produces tails of too heavy fragments seen in the charge distributions. No attempt has been done to tune the model parameters to obtain a better agreement. 

More importantly, multiplicities, kinetic energy spectra and angular distributions for ejecta ranging from protons up to B isotopes have been presented and compared with model predictions to evidence the role of clustering. Such a role results to be clearly evidenced, larger for the heavier system and for the higher bombarding energy as expected considering the increased role of the interaction dynamics.   
At 50~MeV/nucleon the inclusion of the cluster option in AMD produces a variation of up to around a factor 100 in the yield of Be and B isotopes. Accordingly, the use of the cluster option depletes the reservoir of free nucleons and tends on average to produce less excited sources: part of the initial energy is carried out as the kinetic energy of emitted nucleons and clusters. On average, one observes more energetic protons/light clusters in the presence of clustering than without, due to energy and momentum conservation in the NN and N-cluster collisions.
As observed also for the C+C reaction~\cite{TIAN_PhysRevC.95.044613,Tian2_PhysRevC.97.034610}, the inclusion of the clustering improves the model because more IMF with smaller excitation energy are produced in the primary phase.

Our collaboration plans to continue the investigation on cluster effects in medium-light nuclear reactions at intermediate energies with experiments with a larger acceptance in order to better select the various reaction channels and perform more exclusive analyses. This will be pursued using the large acceptance apparatus now at GANIL, where the FAZIA telescopes cover the forward angles complementing the INDRA multidectector. 

\section*{\label{sec:ackn}ACKNOWLEDGMENTS} 

This work was supported by the National Research Foundation of Korea (NRF; Grant No. 2018R1A5A1025563) and partially supported by the IBS grant funded by the Korea government grant number IBS-R031-D1.
This work was also partially funded by the Spanish Ministerio de Economía y Empresa (PGC2018-096994-B-C22). 
The publication was made with the contribution of the researcher C.F. with a research contract co-funded by the European Union - PON Research and Innovation 2014-2020 in accordance with Article 24, paragraph 3a), of Law No. 240 of December 30, 2010, as amended, and Ministerial Decree No. 1062 of August 10, 2021.
We would also like to thank the accelerator staff of LNS laboratories for having provided good-quality beams and support during the experiment.

\bibliography{bibliooc}

\bibliographystyle{apsrev4-2.bst}

\end{document}